\documentclass[10pt,a4paper]{article}
\usepackage{f1000_styles}

\usepackage[numbers]{natbib}

\usepackage{makecell}

\begin{document}
\pagestyle{fancy}

\title{Machine learning based stellar classification with highly sparse photometry data}
\author[1]{Seán Enis Cody}
\author[1]{Sebastian Scher}
\author[2]{Iain McDonald}
\author[2]{Albert Zijlstra}
\author[2]{Emma Alexander}
\author[3]{Nick L.J. Cox}
\affil[1]{Know-Center GmbH, Sandgasse 36, 8010 Graz, Austria}
\affil[2]{Jodrell Bank Centre for Astrophysics, Alan Turing Building, University of Manchester, M13 9PL, UK}
\affil[3]{ACRI-ST, Centre d’Etudes et de Recherche de Grasse (CERGA), 10 Av. Nicolas Copernic, 06130 Grasse, France}

\maketitle
\thispagestyle{fancy}

\begin{abstract}
\paragraph{Background}\ \\
Identifying stars belonging to different classes is vital in order to build up statistical samples of different phases and pathways of stellar evolution. In the era of surveys covering billions of stars, an automated method of identifying these classes becomes necessary.

\paragraph{Methods}\ \\
Many classes of stars are identified based on their emitted spectra. In this paper, we use a combination of the multi-class multi-label Machine Learning (ML) method XGBoost and the PySSED spectral-energy-distribution fitting algorithm to classify stars into nine different classes, based on their photometric data. The classifier is trained on subsets of the SIMBAD database. Particular challenges are the very high sparsity (large fraction of missing values) of the underlying data as well as the high class imbalance. We discuss the different variables available, such as photometric measurements on the one hand, and indirect predictors such as Galactic position on the other hand. 
\paragraph{Results}\ \\
We show the difference in performance when excluding certain variables, and discuss in which contexts which of the variables should be used. Finally, we show that increasing the number of samples of a particular  type of star significantly increases the performance of the model for that particular type, while having little to no impact on other types. The accuracy of the main classifier is $\sim$0.7 with a macro F1 score of 0.61.
\paragraph{Conclusions}\ \\
While the current accuracy of the classifier is not high enough to be reliably used in stellar classification, this work is an initial proof of feasibility for using ML to classify stars based on photometry.
\end{abstract}

\clearpage
\pagestyle{fancy}

\section*{Plain language summary}

Astronomy is at the forefront of the `Big Data' regime, with telescopes collecting increasingly large volumes of data. The tools astronomers use to analyse and draw conclusions from these data need to be able to keep up, with machine learning providing many of the solutions. Being able to classify different astronomical objects by type helps to disentangle the astrophysics making them unique, offering new insights into how the Universe works. Here, we present how machine learning can be used to classify different kinds of stars, in order to augment large databases of the sky. This will allow astronomers to more easily extract the data they need to perform their scientific analyses. 

\section*{Introduction}

The billions of stars surveyed across our Galaxy and beyond by modern telescopes present a broad continuum of different properties. To make sense of this parametric diversity, astronomers typically classify stars as belonging to one or more stellar types (e.g., \cite{MorganKeenan73}). These types often represent a phase in the evolutionary pathway of a particular kind of star, and are defined according to one or more observed properties. Such stellar types can be broad, such as the entire gamut of binary stars, or narrow, such as stars exhibiting a particular type of variability in the emitted light.

The star's spectrum represents one of the primary tools we can use to identify to which classes it belongs. However, the spectra of most stars remain unobserved due to the technical challenges of observing enough light from faint objects. Instead, we rely on photometry, i.e., a measure of the star's brightness in images taken through different ultraviolet, visible and infrared filters. The combination of these different brightnesses at different wavelengths of light define an approximate  spectrum of the star, which is known as a spectral energy distribution (SED). Since the majority of catalogued stars are too faint to be observed with more detailed methods, the SED represents the majority of the information known about most of the stars that have been catalogued so far \cite{stromgren1966spectral}.

If we know the distance to a star and can estimate how much of its light has been absorbed or scattered by interstellar dust, we can correct the SED and estimate the amount of light emitted by the star at the observed wavelengths (e.g., \cite{Fitzpatrick99}). By fitting the corrected SED with atmospheric models of stars, we can estimate the temperature of the stellar surface and, by integrating under the corrected SED, we can determine the luminosity of the star (the rate at which it emits light across all wavelengths). These represent two of the most fundamental properties of a star, which are often used in classification \cite{Hertzsprung1909,Russell1914}. From these data, we can also estimate the stellar radius and, by assuming the star's mass, its surface gravity. However, with only the SED, we are ignorant of many other aspects used to classify stars. We have no information about the star's mass or composition, its movement through space, its detailed spectral features, or any information on the variability of its light, all of which are used in classification (e.g., \cite{Woods11}).

In this respect, the problem of stellar classification relates to similar problems in other domains, such as materials characterisation and Earth imaging, where hyper-spectral remote-sensing observations are used to classify objects in different environments \cite{adep2017exhype,peyghambari2021hyperspectral,BLANCK2017151}.

Modern astronomical catalogues are large: the \emph{Gaia} satellite's Data Release 3 \cite{GaiaDR3} contains approximately 10$^9$ stars with usable distances, to which other catalogues of similar size can be cross-matched to form SEDs. Fitting of these SEDs, as described above, has been automated: in this paper we use the PySSED software package \cite{mcdonaldpyssed} to collect and fit the SEDs of stars observed by \emph{Gaia}. However, an automated stellar classifier that can cope with this wealth of information has yet to be constructed.

In this paper, we use a combination of known stellar positions, distances, motions in the sky, photometry and computed temperature, luminosity, radius and surface gravity as input to a supervised ML multi-label multi-class classification algorithm based on a sample of stars with existing classifications from the SIMBAD  \cite{simbad} astronomical database. We specifically focus on the top-level stellar classes in SIMBAD: binary, emission-line, evolved, low-mass, massive, main-sequence, chemically peculiar and variable stars, and young stellar objects. However, we do not attempt herein to optimise the list of classes using lower-level classes in the SIMBAD hierarchy.

The principal feasibility of using ML methods to classify stellar objects has been shown in \cite{gabruseva2020}, albeit on synthetically generated data. A number of works (e.g. \cite{clarke2020identifying,cunha2022photometric,chaini2023photometric,zeraatgari2024machine}) have used an ML approach to classify point sources into galaxies, quasars, and stars. ML has also been demonstrated in differentiating specific types of star, such as planetary nebulae \cite{Iskander20}, symbiotic stars \cite{jia2023identifying}, variable stars \cite{naul2018recurrent,pantoja2022semi}, and exoplanets in orbit around stars \cite{Hayes20}. In contrast to these works, we focus on the entire range of stellar types, but only on stars specifically, and on the problem of how to classify them into different stellar classes. Stellar classification has been succesfully carried out for hundreds of millions of stars from spectra (e.g. \cite{GaiaDR3}).  We focus on the challenge of classifying the much larger number of stars for which spectral observations are impractical or impossible, and for which only SEDs can be obtained.

To sum up, the main objectives of our paper are (1) to demonstrate the applicability of ML in classification of the entire range of stellar SEDs, given only sparse photometric measurements based on a random sample of the SIMBAD database; (2) to measure how inherent sampling bias and associated class imbalance in astronomical literature affects the classifier's performance; and (3) assess how limitations caused by minority classes in the training database can be improved by the acquisition of more data.

\subsection*{Challenges}

There are several key challenges in this work that differentiate it from more typical supervised-learning ML problems. Firstly, the data used are a collection of many different sky surveys: these surveys cover different regions of the sky at differing sensitivities, and some stars bright enough for one survey may be too faint for another or, conversely, some stars may be observable by one survey but saturated in another. Consequently, a large fraction of data values are not present, i.e., the dataset is very sparse. This makes it much harder to train accurate ML models on the data.

Secondly, the stars for which classifications have been produced are severely biased. Some regions of the sky have been specifically targeted for particular kinds of star (e.g., the footprints of surveys for variable stars), thus leading to a spatial bias in the recovery of certain stellar types. Furthermore, certain types of stars are globally better classified than others, e.g., large lists of massive stars have been identified as these stars are bright and well-studied, while there are very few classified main-sequence stars, since these represent the majority of stars and are therefore often considered unworthy of classification. This strong bias makes both training and evaluating the ML models challenging.

Thirdly, stars may fall into multiple categories, though certain category combinations are prohibited: e.g., a star may be both massive and variable, but cannot be both massive and low-mass. This problem adds complexity to producing the output of a ML algorithm and defining its success.

Finally, as mentioned above, specific stellar classes are identified based on data types other than those we have collected (e.g., temporal variability or characteristics of detailed spectra). Therefore, it is \emph{a priori} not clear how well stars can be classified based on these data alone. This fundamentally differentiates our problem from certain other supervised ML problems. For example, in many image-classification problems, such as the widely used Imagenet dataset \cite{ILSVRC15}, humans themselves can do the classification given the images. While this does not necessarily mean that is possible in practice to manage to train an ML algorithm to do the same task, it is at least known that such problem is theoretically solvable, given a good enough training algorithm and sufficient training samples. This is not the case for our problem setting, and we thus have no \emph{a priori} knowledge of how well, even with unlimited training samples, a classification can actually be made for certain stellar classes.

\section*{Methods}
In this section we describe the data used, the PySSED spectral energy distribution fitting algorithm, the pre-processing steps, machine-learning methods and tuning procedure.

\subsection*{PySSED}
PySSED is a Python code for fitting stellar spectral energy distributions. Given a star, list of stars, or region of the sky, it will retrieve the star's co-ordinates from SIMBAD, a database for astronomical objects outside of the Solar System \cite{simbad}, and perform a search using the associated VizieR database\footnote{https://vizier.u-strasbg.fr/viz-bin/VizieR} for multi-wavelength photometric brightness measurements, distances and other requested parameters. PySSED will then correct the photometry for interstellar reddening at the location and distance of the star, before fitting the corrected photometry with a stellar model atmosphere to estimate the temperature and luminosity of the star. Differences between the observed and modelled fluxes are also recorded. Note that the interstellar reddening correction is based on \emph{Gaia} spectroscopic data, meaning \emph{Gaia}-derived stellar parameters are no longer a completely independent comparison. An in-depth analysis of the SED fitting outputs, procedures, quantification of the impact of low photometric information, and a discussion of data biases can be found in \citet{mcdonaldpyssed}, while \citet{PySSED_use} shows an example where PySSED is used on \emph{James Webb Space Telescope} data.

\subsection*{Data}

Our training set of stellar classifications comes from SIMBAD. While SIMBAD compares more objects than only stars, in this study we only look at stars, and specifically at the following nine stellar classes: (1.6) binary stars, (1.10) emission-line stars, (1.4) evolved stars, (1.8) low-mass stars, (1.1) massive stars, (1.3) main-sequence stars, (1.5) chemically peculiar stars, (1.9) variable stars and (1.2) young stellar objects (YSOs). The numbers in parentheses refer to the taxonomy used in SIMBAD. The stellar classes and the abbreviations that will be used throughout the paper are summarized in Table \ref{table:abbreviations}.

\begin{table}
\centering
\caption{\label{table:abbreviations}Star types used in the study and their abbreviations. }
\begin{tabular}{c|c}
    Star type & Abbreviation  \\
    \hline
    Binary or Double Star & ** \\
    \hline
    Emission Line Star & em* \\
    \hline
    Evolved Star & ev* \\
    \hline
    Low-Mass Star & lm* \\
    \hline
    Massive Star & ma* \\
    \hline
    Main Sequence Star & ms* \\
    \hline
    Chemically Peculiar Star & pe* \\
    \hline
    Variable Star & v* \\
    \hline
    Young Stellar Object & y*o\\
    \hline
\end{tabular}
\end{table}
These stellar classes are not exclusive, thus there is overlap expected between certain classes. For example, many evolved stars are variable. (However, between other classes there can be no overlap, e.g. stars can't be both massive and low-mass). Allowed and not-allowed overlaps are detailed in Table \ref{tab:overlap}.

\begin{table}
\caption{Allowed overlaps between the nine star types.}\label{tab:overlap}
\centering
\begin{tabular}{l|ccccccccc}
\hline\hline
Category   & Binary & Emission & Evolved & Low-mass & Massive & Main-seq. & Chem.~pec. & Variable & Young\\
\hline
Binary     &---& Y & ? & Y & Y & Y & Y & Y & Y \\
Emission   & Y &---& Y & ? & Y & ? & Y & Y & Y \\
Evolved    & ? & Y &---& N & Y & N & Y & Y & N \\
Low-mass   & Y & ? & N &---& N & Y & ? & Y & Y \\
Massive    & Y & Y & Y & N &---& Y & ? & Y & Y \\
Main seq.  & Y & ? & N & Y & Y &---& Y & Y & N \\
Chem.~pec. & Y & Y & Y & ? & ? & Y &---& Y & Y \\
Variable   & Y & Y & Y & Y & Y & Y & Y &---& Y \\
Young      & Y & Y & N & Y & Y & N & Y & Y &---\\
\hline
References & \cite{chen2023binary} & \cite{kogure2010astrophysics} & \cite{herwig2005evolution} & \cite{luhman2012formation} & \cite{motte2018high,eldridge2022new} & \cite{schwarzschild2015structure} & \cite{preston1974chemically} & \cite{conroy2018complete} & \cite{koenig2014classification} \\
\hline
\multicolumn{10}{p\textwidth}{Y = overlap expected, ? = some overlap possible, N = overlap disallowed.}\\
\hline
\end{tabular}
\end{table}

SIMBAD contains over 1 million stars of the classes we use. In this study, we use two subsets of those stars for analysis with PySSED. The first is a subset of 10\% randomly sampled stars, resulting in 117\,128 stars. This choice was largely dictated by computational time, originating from evaluating multiple queries per star at the CDS SIMBAD database during the PySSED preprocessing steps. This will be referred to as the "main dataset".
In the main dataset, the most prevalent star type are variable stars, followed by evolved stars and binary stars, while the least prevalent class - only 0.24\% -- are massive stars (see Fig.~\ref{fig:sparsity}a).
While we use random sampling of the SIMBAD database, this does not correspond to a random sample of stars in reality, as the stars in SIMBAD are not a random sample of real stars. Therefore, as a second dataset, we extended the main dataset with 80,000 additional single label main-sequence stars. Main-sequence stars are only a very small fraction of the stars in the random dataset, but are severely underrepresented in SIMBAD: in reality they represent the majority of stars on the sky. We will refer to this dataset as the "extended dataset". Comparing the results between the two datasets  allows us to show how adding more samples of a specific star type  affects the performance of the classifier.

\begin{figure}
	\centering
	\includegraphics[width=0.5\textwidth]{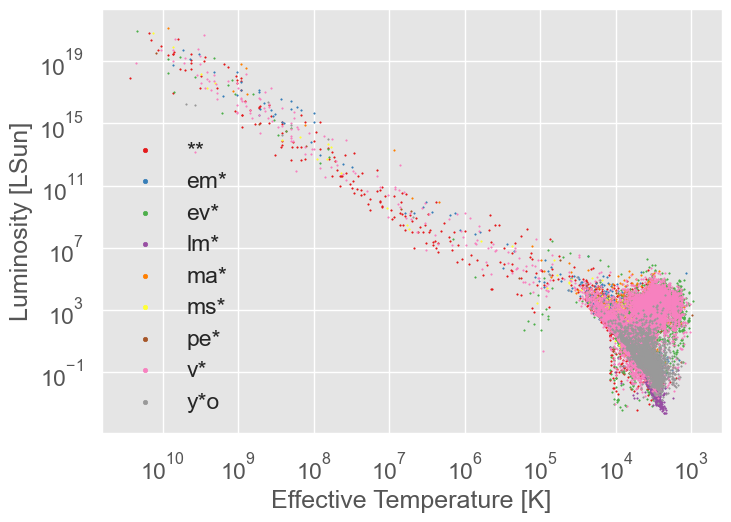}\includegraphics[width=0.5\textwidth]{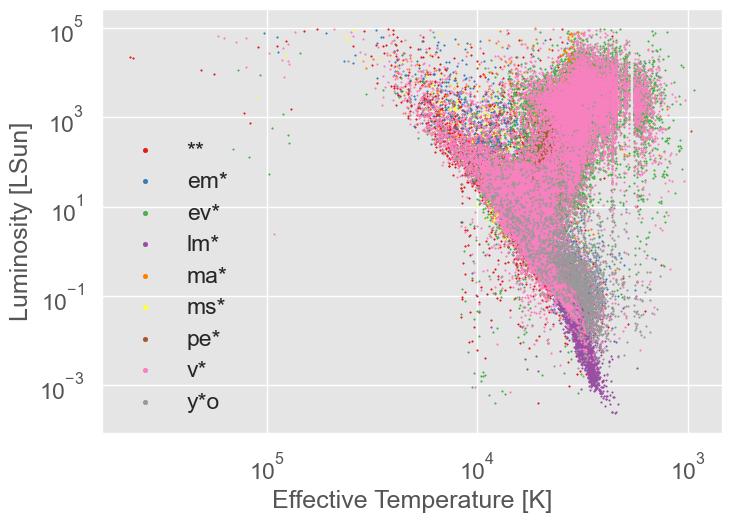}
	\caption{\label{fig:HR}Hertzsprung--Russell diagram of the stars in the dataset used, with effective temperature and luminosity estimated with PySSED. The right panel is a zoom in of the left plot. The different star types are indicated by color.}
\end{figure}

\begin{figure}
\centering
a)\includegraphics[width=0.3\textwidth]{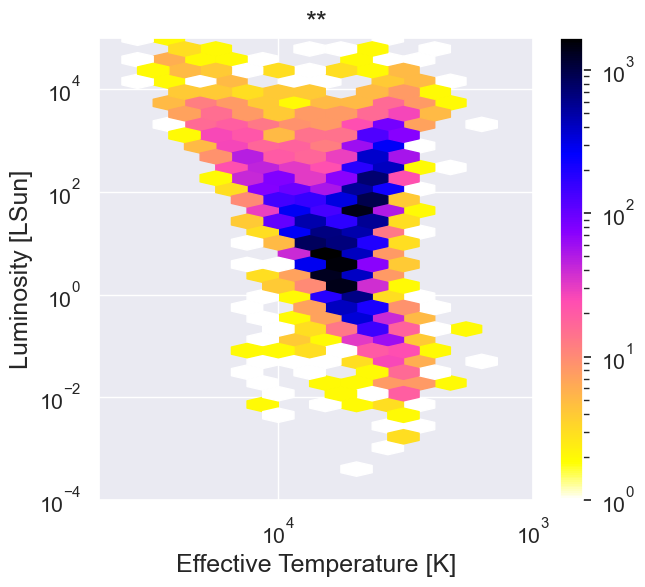}
b)\includegraphics[width=0.3\textwidth]{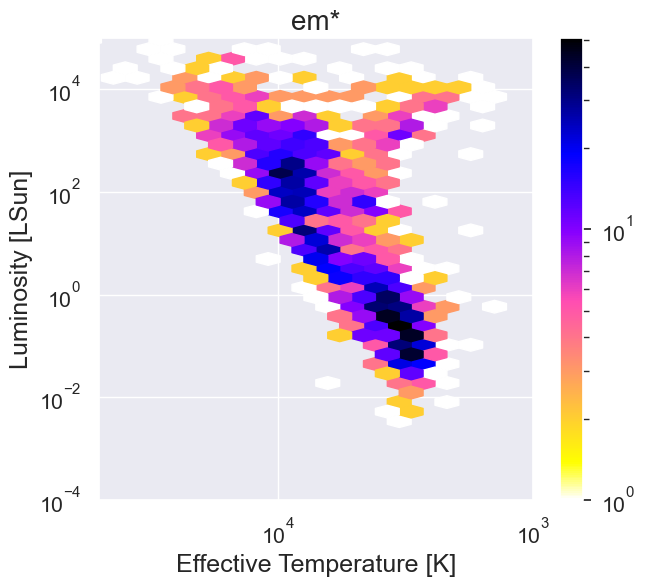}
c)\includegraphics[width=0.3\textwidth]{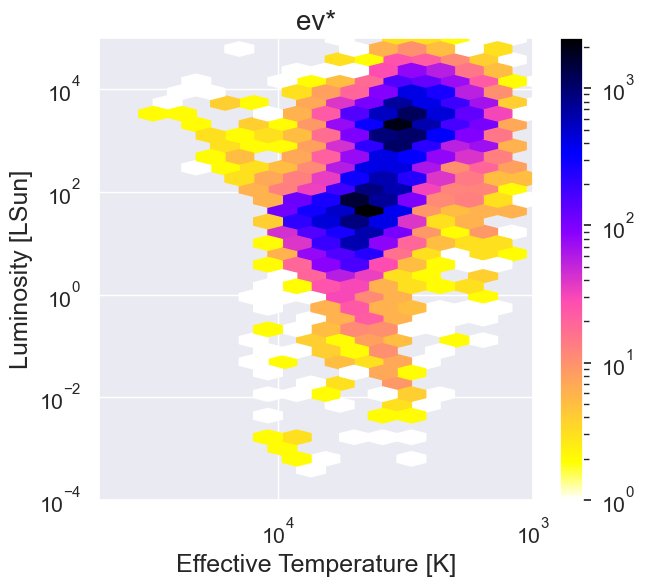}
d)\includegraphics[width=0.3\textwidth]{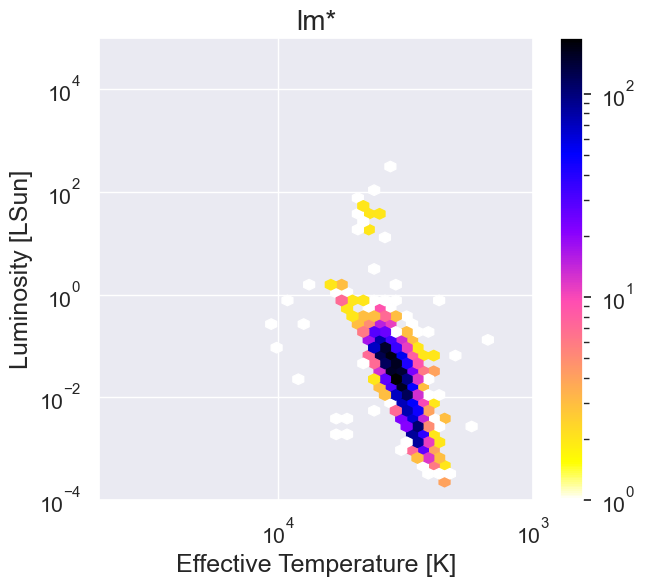}
e)\includegraphics[width=0.3\textwidth]{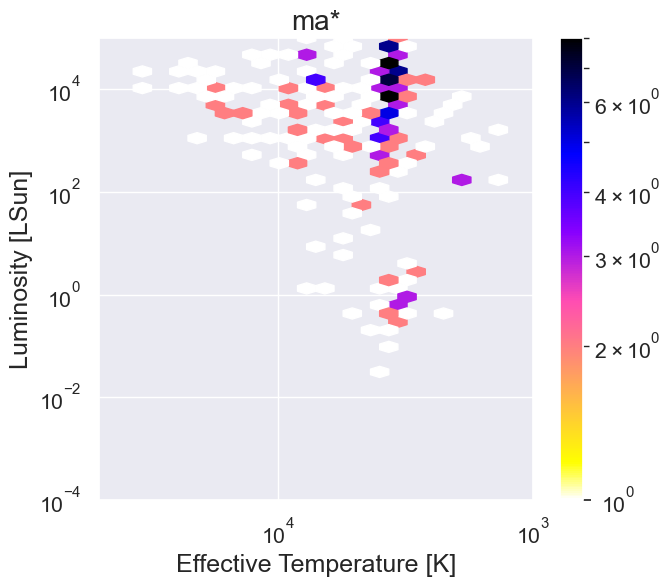}
f)\includegraphics[width=0.3\textwidth]{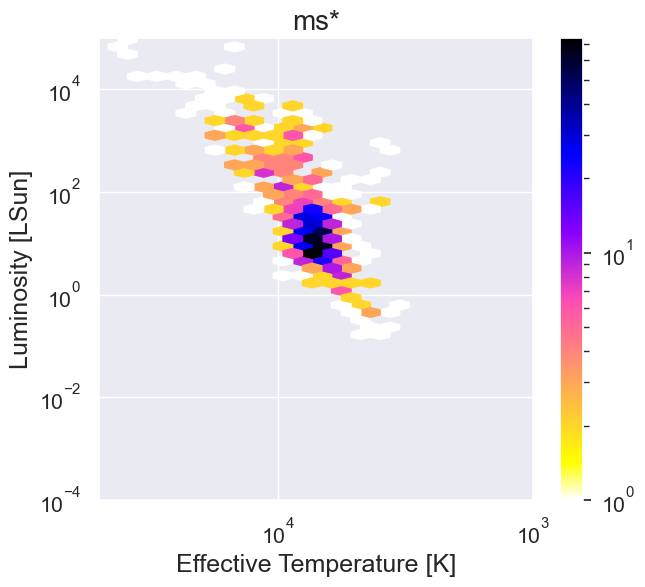}
g)\includegraphics[width=0.3\textwidth]{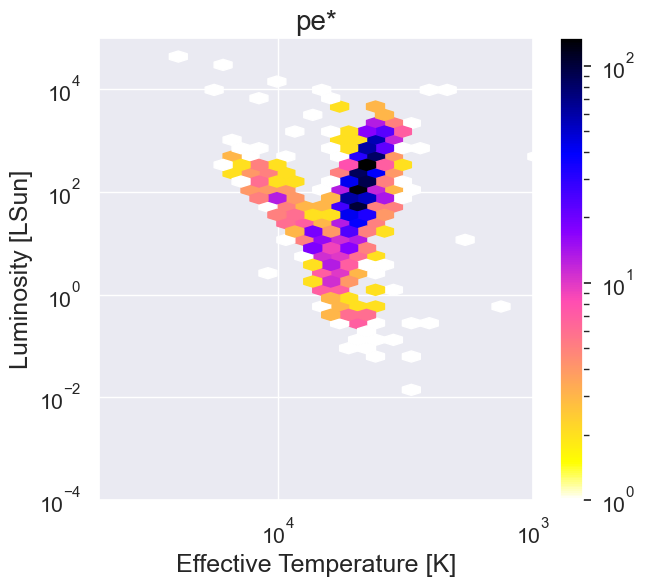}
h)\includegraphics[width=0.3\textwidth]{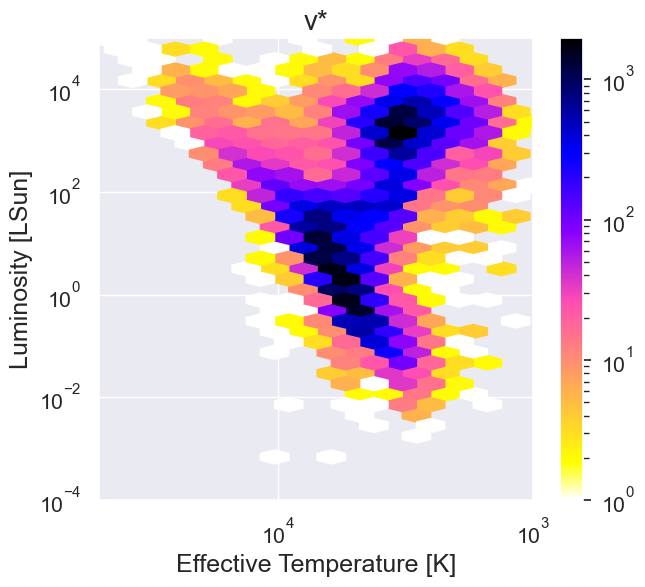}
i)\includegraphics[width=0.3\textwidth]{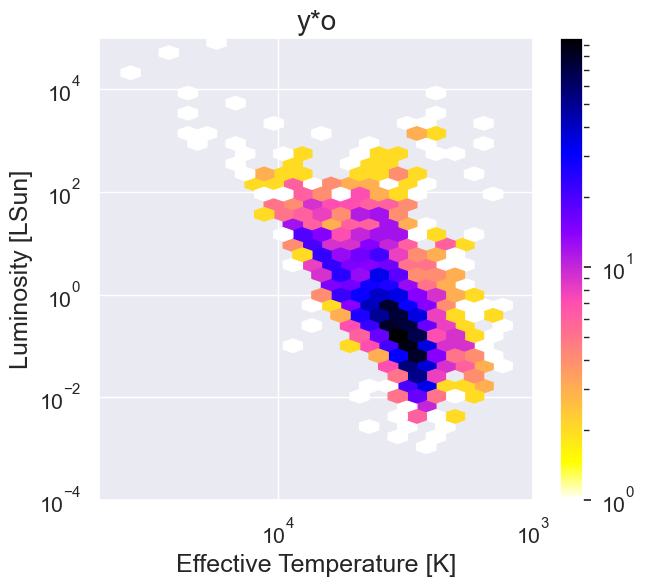}
j) \includegraphics[width=0.3\textwidth]{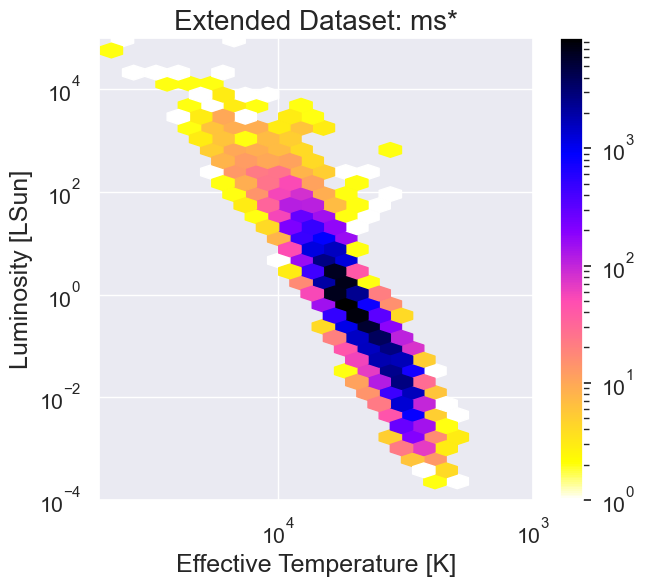}
 \caption{\label{fig:HR_classes}Density Plot HR Diagrams By Class Association}
\end{figure}

The results of fitting our "main dataset" with PySSED is shown in Fig.~\ref{fig:HR}. Several deficiencies in the PySSED output data are apparent. Most notable among these are the set of stars stretching to unphysically high temperatures and luminosities. This is the most significant source of error in the PySSED analysis of these stars, and is caused when hot stars are over-corrected for interstellar reddening, often due to over-estimated distances or inaccuracies in the coarsely sampled interstellar-reddening maps. Removal of these stars did not significantly improve the ML classifiers. Therefore, effective temperature and luminosity output by PySSED were directly used as input features for the ML models. Other sources of error, including luminosity changes from errors in distance or deficiencies in individual photometric data, are small by comparison. The H--R diagram for each class is shown in Fig.~\ref{fig:HR_classes}.

\subsubsection*{Train test split}
ML algorithms tend to have a large number of parameters and can therefore easily overfit on the training data. This means that when using the trained model to make predictions on the training data itself the predictions are very good, but that model does not work well on unseen data. The scores on the training set can thus be misleading. Therefore, it is essential to withhold data from training that can then be used for an independent evaluation of the algorithm. This is usually referred to as the "test" set. Additionally, if one performs model tuning -- which is essentially trying out multiple algorithms -- one needs a third set that is used as evaluation set in the tuning phase, or alternatively use cross-validation on the training set. Cross-validation provides a more robust and accurate measure of performance during tuning, and we therefore opted for 5-Fold Cross Validation (CV).
Deciding how to split up a dataset into training and test set is not trivial. Often, random sampling is employed. In our case standard random sampling is not feasible due to the high class imbalance, as the number of samples for the minority classes in the test set would be too low for getting robust results. Therefore, we used a modified random sampling that ensures that a minimum number of each star type is contained in the test set. To that end, at least 40 instances, accounting for about 15 percent of the minority class, were included per class in the test set used for final evaluation, before randomly sampling from the remaining data. We did this independently for the main and for the extended dataset. All scores reported in this paper are on the respective independent "test" datasets. Furthermore, in accordance with avoiding nonsensical predictions, if no class is predicted in a multi-label scenario, the class of highest confidence is selected as the label prediction.

\subsection*{Preprocessing}
The default output of PySSED contains both the original photometry and the reddening-corrected photometry, the photometry of the model fit, the temperature, luminosity, sky position and a number of other parameters ("features" in ML-jargon), resulting in 442 feature columns. Description of the default outputs and bands used are given in \citet{mcdonaldpyssed}. Some of these can be reasonably grouped together, which we do as part of data preprocessing, to reduce the data dimensionality.

The first steps taken in preprocessing the data involved explicitly setting any missing values (which differed in representation) to NaN (not-a-number) and converting the Earth-centric location variables, RA (right ascension) and Dec (declination) to Galactic coordinates.

The largest feature engineering step conducted during preprocessing involves condensing the information contained in the photometry into a single data feature at each wavelength. To do this, we make use of the observed photometric flux in each filter ($F_{\rm o}$), the error on that flux ($\epsilon$), and the modelled flux in the same filter ($F_{\rm m}$). We then calculate an uncertainty, $\sigma$, via

\begin{equation}
    \sigma = \sqrt{\epsilon^2 + \left(\frac{F_{\rm o}}{10}\right)^2}
\end{equation}

This artificial addition in quadrature of 10\% of the observed flux sets an uncertainty floor, which accounts for systematic issues with the observations. These can include uncertainty in the zero-point calibration of the survey, or differences between the assumed and real photometric filter transmission curves. This allows us to generate a significance ($S$) for each filter, representing how deviant the observation is from the model:

\begin{equation}
    S = \frac{\log(F_{o} / F_{m})}{\log(\sigma + 1)} 
\end{equation}

The use of logarithms accounts for the typical uncertainties on astronomical observations being given in magnitudes, and the tendency of observations to be close to the photon limit of detectors, thus have asymmetrical errors due to Poisson noise. It also prevents the most extreme deviations from dominating the feature space. By taking the ratio of observed to modelled flux, we also remove the dominant effects of stellar temperature, which is the most significant determinant of the shape of a stellar SED.

\begin{figure}
	\centering
	a)\includegraphics[width=0.45\textwidth]{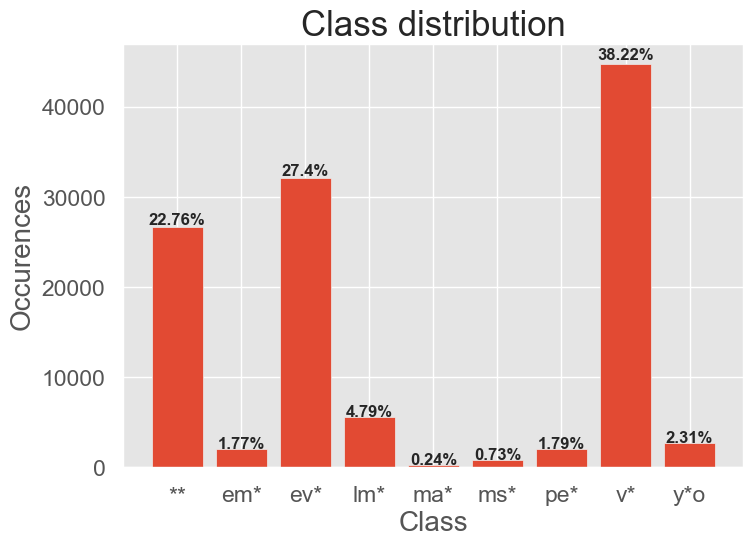}
 b)\includegraphics[width=0.45\textwidth]{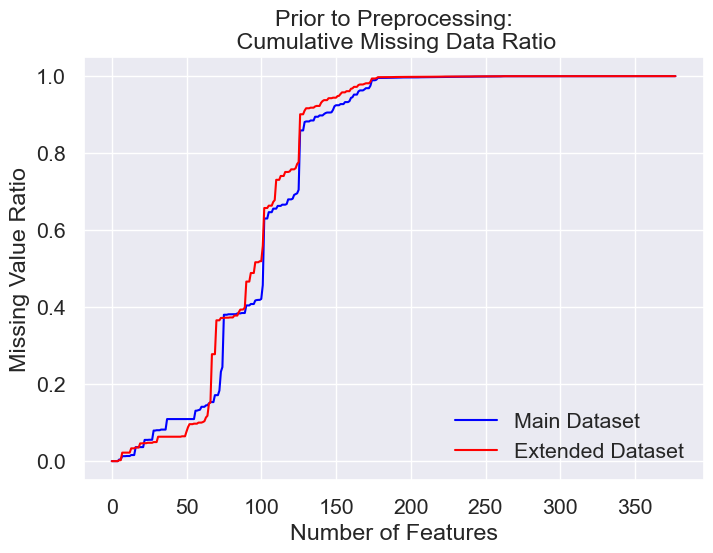} c)\includegraphics[width=0.45\textwidth]{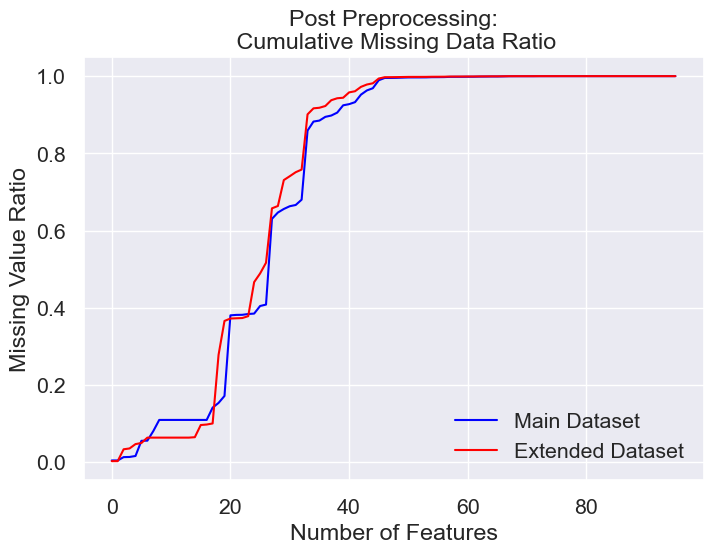}
	\caption{\label{fig:sparsity} a) Class imbalance (distribution of star types) in the main dataset - multi-label samples are evaluated separately as single class instances for each of their class associations b) cumulative sparsity of the main dataset before preprocessing is applied, c) cumulative sparsity of the main dataset after preprocessing is done.}
\end{figure}

After applying the above mentioned preprocessing steps to the default PySSED output data, 96 features are retained. Still, for individual stars, a lack of observations means data are not necessarily available for all of those features. A quantitative analysis of the sparsity of the main dataset is shown in Fig.~\ref{fig:sparsity}b and Fig.~\ref{fig:sparsity}c, which depict the 74\% total missing data in the main dataset. When regarding Fig.~\ref{fig:sparsity}c for example, we find that only 27 out of the 96 features contain values for at least 50 percent of samples, i.e., 69 features have missing values for more than 50 percent of samples. There are various reasons for these missing data, though two factors dominate. First, the dynamic range to bright (saturated) and faint (missing) stars varies from survey to survey, and no survey samples the entire range of stellar brightnesses; some surveys were particularly chosen because they cover a small number of very bright stars that saturate in other surveys. Second, the spatial coverage of many surveys is limited, either because the telescope used had a small field of view and performed targeted observations, rather than covering the entire sky, or because the telescope's field-of-view is limited by the hemisphere it is in.

\subsection*{Feature packages}
The features in the SIMBAD data can be separated into several distinct groups, with different physical and conceptual meaning. 

Some of the features, such as the photometry, temperature and luminosity, have direct physical meaning for stellar classification.  In contrast, some other features, such as distance or Galactic position also do have a physical meaning, but only indirectly in the context of classifying stars. Different types of star are unequally distributed across the Galaxy, both in reality (e.g., young stars are found most frequently near the Galactic Plane) and in our dataset (e.g., the Galactic Bulge and Magellanic Clouds have been comparatively well-sampled for variable stars by microlensing surveys). Thus, position actually has predictive value for classifying stars, and a good ML training algorithm should be able to exploit this. However, the predictive skill -- the performance of the classifier measured by metrics such as accuracy and f1-score --  that comes from those variables is not based on the physics of the star, and it is thus debatable whether one should actually use those variables, even when they can increase the skill of the classifier. We will refer to the stellar position, proper motion across the sky and distance as "bias features" throughout this paper. In practice, the answer to whether one should or should not use those bias variables will depend on for what exact use case the classifier shall be used. If the goal is to make the best possible guess of a star type, given all information available, then the bias variables should be used. However, if the task is to infer how much we can tell about the type of the star from photometric measurements, then the bias variables should be explicitly excluded.

\begin{figure}
	\centering
\includegraphics[width=0.5\textwidth]{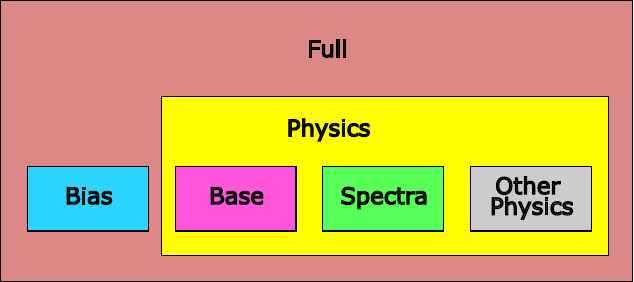}
	\caption{\label{fig:feature-pckages} Overview of feature packages. There are four feature groups (Bias, Base, Spectra and Other Physics) and five feature packages built from them (Bias, Base, Spectra, Physics and Full).}
\end{figure}

We grouped the variables into four categories: variables referring to position ("bias variables"), variables describing photometric measurements ("spectra variables"), effective temperature and luminosity as output by PySSED ("base variables") and a final group containing physics variables that are not direct photometric measurements (namely: the metallicity and surface gravity PySSED adopts for the star, the interstellar extinction, and any measurement of spectroscopic temperature). The  feature categories are summarized in Table \ref{tab:FeatureSets}. From these four groups, we form five different feature packages: "Bias", consisting only of the bias variables; "Base", consisting only of the two base variables; "Spectra", consisting only of the spectra variables; "Physics", consisting of the spectra plus the other physics variables; and "Full", consisting of all variables. This is visualized in Fig.~\ref{fig:feature-pckages}.

\begin{table}
    \caption{Feature Groups}
    \label{tab:FeatureSets}
\centering
\begin{tabular}{p{3cm}|p{13cm}}
\hline\hline
Feature Set & Features\\
\hline
Base Variables& Effective Temperature (K), Luminosity (LSun)\\ 
Bias Variables & Right Ascension (degrees), Declination (degrees), Proper Motion Right Ascension (mas/year), Proper Motion Declination (mas/year), Distance (pc)\\
Spectra Variables & Transformed Spectra [85 features (Jy)]\\
Physics Variables & Base Variables, Spectra Variables,\newline Other Physics [Metallicity (dex), Surface Gravity (dex), Extinction (mag), Spectroscopic Temperature (K)]\\
\hline
\end{tabular}
\end{table}

\subsection*{Machine-learning methods}
The problem considered here is a supervised learning classification problem with continuous input features. Herein, models are tasked with either explicitly or implicitly learning some approximation of the sample data's class distribution. A common approach is to have models learn decision boundaries with the aim of separating the support sets of the class distributions, which ideally are aimed at generalizing from the sample to the population. In principle, there is a wide range of supervised learning methods for classification tasks available, e.g. logistic regression, deep neural networks, support vector machines. However, the high sparsity of the data strongly restricts which ML methods can be used. Crucially, it renders the naive approach of simply completely dropping all features with missing values, to generate a subset with no missing values impossible. As this is not possible in our case, there are two approaches left: (1) use data-imputation techniques, or (2) use ML techniques that can natively deal with missing values. Data imputation refers to techniques that fill up missing values with plausible values. There is a wide range of data imputation techniques in the literature \cite{lin2020missing}. Initial tests that we carried out with simple imputation techniques and subsequently training various ML models on them did not yield promising results. Data imputation can easily generate data that is not physically meaningful. Therefore, we settled on approach (2), using a supervised classification algorithm that can directly deal with missing data, more specifically the XGBoost algorithm \cite{chen2016xgboost}. XGBoost is a widely used tree-based supervised ML algorithm for multi-class classification tasks. It is especially well-suited to tabular data, where it usually outperforms deep-learning algorithms \cite{shwartz2022tabular}, and can natively deal with missing values. We thus opted to solely look at XGBoost models in detail for our problem. XGBoost -- like most other ML training algorithms -- has a large number of so-called hyperparameters, in addition to the parameters that are learned by fitting on the training data. These need to chosen in advance, before commencing training. Hyperparameters of XGBoost are, amongst others, the learning rate and the maximum depth of each tree. The most common approach to choosing suitable hyperparameters is to perform systematic model tuning. For this, the model is trained with different hyperparameters on the training set, and then evaluated on the validation set, and subsequently the hyperparameter configuration that yielded the best performance on the validation set is used. Alternatively, as we have done, 5-Fold Cross Validation (CV) can be used to more accurately estimate algorithm performance, in place of a classical validation set.

The tuning procedure was performed individually for each of the five feature packages, resulting in five hyperparameter configurations. For tuning, we used Bayesian optimization as implemented in the Optuna Python library \cite{optuna_2019}, utilizing the Tree-Structured Parzen Estimator (TPE) \cite{bergstra2011TPE} for parameter search coupled with median pruning to limit the search space. The tuning grid -- that is, the hyperparameter space that is explored in the tuning process -- is shown in Table \ref{tab:tuning_grid}. The introduction and tuning of nine class weight parameters, to influence sample weights, was explicitly added to counteract the high class imbalance in the data. The class weights were additively combined per multi-label sample and used to directly influence the gradient computation in each boosting step, by scaling the sample's impact on the loss function. The resulting hyperparameter configurations are contained in the code repository accompanying this paper.

\begin{table}
    \caption{Grid for Hyperparameter Tuning XGBoost via Baysian Optimization.}
    \label{tab:tuning_grid}
    \centering
    \begin{tabular}{ c | c | c | c }
        \hline
        Hyperparameter & High & Low & Step Size \\ [0.5ex] 
        \hline\hline
        Class Weights 0-8 & 1.0 & 0.001 & 0.001\\
        \hline
        max\_depth & 10 & 3 & 1\\
        \hline
        min\_child\_weight & 20 & 0 & 2\\
        \hline
        gamma & 10.0 & 0.0 & 0.2\\
        \hline
        subsample & 1.0 & 0.2 & 0.1\\
        \hline
        colsample\_bytree & 1.0 & 0.3 & 0.1\\
        \hline
    \end{tabular}
\end{table}

\subsection*{Evaluation metrics}

Evaluating the predictions of a multi-class, multi-label classifier on highly imbalanced data is not trivial, and care has to be taken to choose the right evaluation metrics. Uncorrected accuracy (the fraction of correctly classified stars) can be highly misleading for unbalanced datasets, and can give incorrect impressions of good predictions. For multi-label predictions -- where one sample can have multiple labels simultaneously -- an additional complication is that it is not trivial to decide which predictions should be counted as correct and which not. E.g., if a star that is both a binary and a main-sequence star, but the classification model only predicts binary - should we count it as a correct or an incorrect prediction?

Some standard choices we choose to visualize are the raw accuracy and F1 score. Of these demonstrated metrics, accuracy is included purely for completeness sake, as we deem it inadequate in representing the performance on a multi-label dataset with extreme class imbalance. After all, the standard definition of accuracy requires a perfect match, i.e., even if three out of four star types were predicted correctly for a sample, this would result in a classification score of 0 for that sample, masking the fact that the model prediction was almost completely correct. Depending on problem context this can be a highly significant issue. Furthermore, with extreme class imbalance, as shown in Fig.~\ref{fig:sparsity}a, a very high accuracy of about 88.38 can be achieved whilst predicting only the three majority classes correctly, hiding the fact that six out of nine stellar classes would never be predicted correctly.

\begin{align}
    &\mathrm{Accuracy} = \sum_{n=1}^{number\_of\_samples}\frac{\mathrm{ \mathbb{1}(predicted\_label == true\_label)}}{\mathrm{number\_of\_samples}}
 \label{eq:accuracy}
\end{align}

A common way to evaluate highly imbalanced classification tasks is to use Precision and Recall, which are defined via True Positives (TPs), False Positives (FPs) and False Negatives (FNs):

\begin{align}
    &\mathrm{Precision} = \frac{\mathrm{TPs}}{\mathrm{TPs + FPs}},\\[3ex]
    &\mathrm{Recall} = \frac{\mathrm{TPs}}{\mathrm{TPs + FNs}}
 \label{eq:prec-rec}
\end{align}
and their harmonic mean, the F1-score
\begin{align}
    &\mathrm{F1} = 2 \times \frac{\mathrm{recall} \times \mathrm{precision}}{\mathrm{recall + precision}} .\label{F1_score_equation}
\end{align}

The precision score evaluates, in what percentage of cases, the algorithm’s predicted class was truly the correct class. Recall, on the other hand, measures the percentage of stars of a given class that are not predicted as that class. In other words, a bad recall value indicates that the model tends to not detect the class label when confronted with a sample of that type. The F1 score is the harmonic mean of precision and recall. 

The three scores are, however, defined for binary classification problems (problems where an object can be only one of two classes). Extending this to multi-class  single-label classification -- where there are multiple classes, but a star can only belong to one single class -- can be done by computing the three scores separately for each class. In our case, we would thus get precision, recall and F1 for binary stars, precision, recall and F1 for main sequence stars, and so on.
 However, in our case, a single star can belong to multiple classes, and TP, FP and FN are thus ambiguous and need to be defined in more detail. For this, two possible approaches exist: 
\begin{enumerate}
    \item  a perfect matching is performed, where every class a star belongs to must be predicted correctly. Thus, if a star belongs to one single class, then a TP means that only that class, and no other additional class was predicted. 
    \item  a partial matching is calculated, where every class a star belongs to is seen as a separate entity. For example, if a single test star is both binary and main-sequence, but is predicted as both binary and low-mass, it is a TP for precision, recall and F1 for binary stars, a FP for precision, recall and F1 for low-mass stars, and a FN for precision, recall and F1 for main sequence stars.

In this paper, we have chosen option (2).
\end{enumerate}

As mentioned, the F1 score is defined on a by-class basis, meaning that in our case we get nine F1 scores, one for each stellar class. There are several ways of extending the F1 score in order to get a single value over all classes.

The simplest way is the \textbf{micro}-averaged F1 score. Here, all TPs, TNs and FNs over all classes are aggregated, and the F1 score computed from these aggregated values. The micro-averaged F1 score suffers from similar deficiencies in regards to class imbalance as the raw accuracy metric. Class labels are disregarded and the balance between precision and recall is evaluated for the entire test set.

\textbf{Macro}-averaging, on the other hand, calculates the F1 score for each class and averages the results via arithmetic mean. This is our benchmark metric of choice, as the metric is harsh on misclassification of individual classes. In other words, each class contributes equally to the final score.

\textbf{Weighted} averaging is based on a similar notion, however applies weights to the class-specific F1 scores, equal to their number of instances, prior to averaging. Thereby the imbalance is further exaggerated, making the metric less telling, for datasets of high class imbalance, compared to the macro approach, where all classes are equally important.

Finally, the \textbf{samples} average applies the unweighted arithmetic mean to the F1 score per sample in the test dataset. This is advantageous when dealing with a high amount of multi-label instances, however when the multi-label count is low, it will be close to the accuracy score we initially disregarded, suffering in part from similar drawbacks.

To get a complete picture, in this paper we use all four versions of the F1 score in parallel.

\section*{Results}

\subsection*{``Main'' dataset}

\begin{figure}[h!]
	\centering
a)\includegraphics[width=0.45\textwidth]{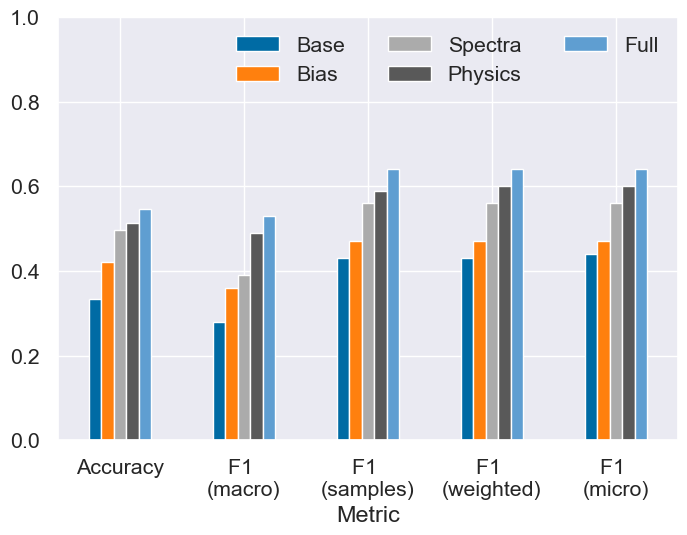}b)\includegraphics[width=0.45\textwidth]{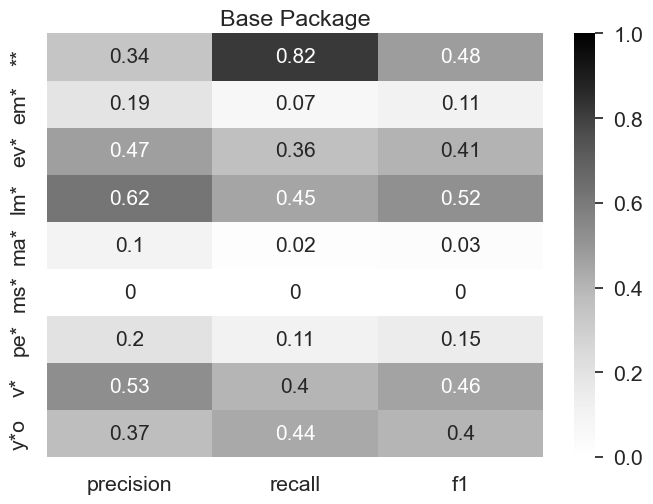}
c)\includegraphics[width=0.45\textwidth]{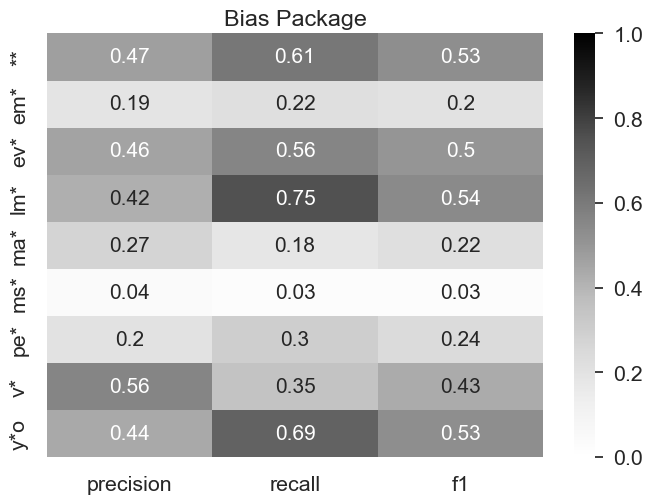}d)\includegraphics[width=0.45\textwidth]{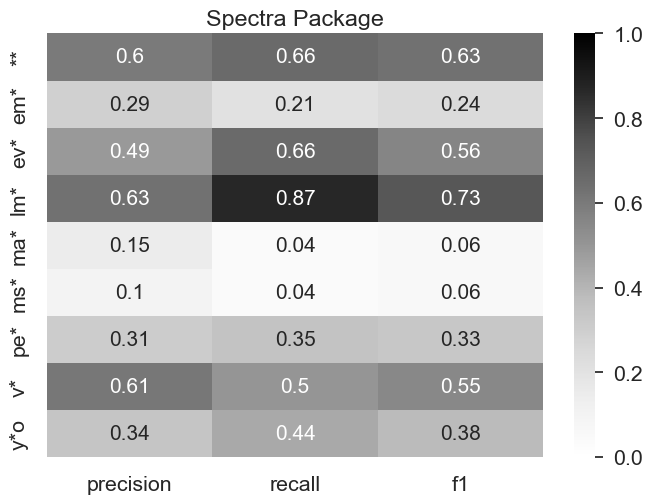}
e)\includegraphics[width=0.45\textwidth]{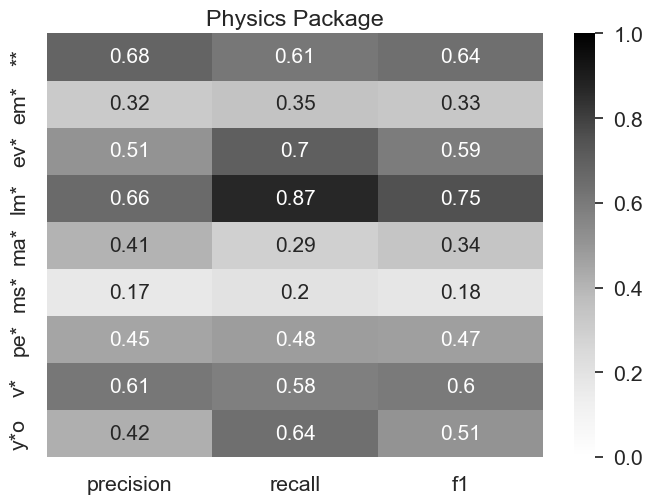}f)\includegraphics[width=0.45\textwidth]{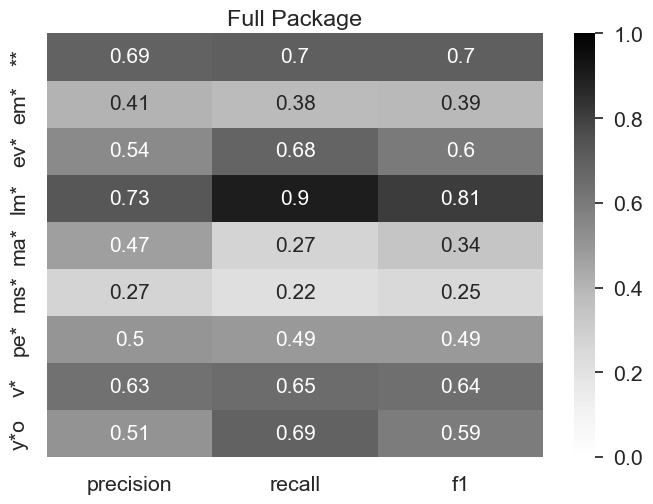}
	    \centering
    \begin{tabular}{c|c|c|c|c|c|c|c|c|c}
        \hline
        \multicolumn{10}{c}{Test set sample counts per class.}\\
        \hline\hline
         & ** & em* & ev* & lm* & ma* & ms* & pe* & v* & y*o \\
        \hline\hline
        Samples & 5345 & 467 & 6345 & 1139 & 100 & 198 & 460 & 9410 & 564\\
        \hline
    \end{tabular}
    \caption{\label{fig:main-res} Accuracy on the main (random 10\%) dataset for models trained with the five different feature packages. Panel (a) shows the results aggregated over all star types. Panels (b-f) show the results for individual star types.}

\end{figure}

We start by analyzing the skill of the classifier trained on the main dataset. The results are shown in Fig.~\ref{fig:main-res}. Panel (a) shows the results aggregated over all stellar classes, with the metrics as described above in the ``Metrics'' section. While the absolute values of the metrics are somewhat different, there is a clear and consistent ordering in the skill of the models in terms of which feature package they use, independent of which metric is considered. The Base model -- the one only using the effective temperature and luminosity from PySSED as predictors -- has lowest skill, while the Full model has highest skill. The Physics model is second-best, followed by the Spectra model and the Bias model. That the Full model (the one using all available features after preprocessing) has highest skill is not surprising. The same is true for the fact that the Physics model is better than the Base model, as the Base features are a subset of the Physics features. The rest of the ordering of the models does however provide some valuable insights. That the Bias model provides higher skill than the Base model may seem intriguing, as there is no physical meaning in the Bias features; however, as discussed in the Introduction, it is debatable whether the predictive value of the Bias features should actually be used, as it represents a combination of the (real) Galactic distribution of stars and their (artifical) inhomogeneous sampling in the dataset. With the same reasoning, the skill of the Full model needs to be treated with caution, as the only difference between the Full and the Physics model is the inclusion of the Bias features. Whether this skill increase is actually valid will depend on whether one considers the predictive value of the Bias features valid, which in turn will depend on the use case the model would be used for.

More promisingly, the Spectra model is clearly better than the Base model. The Spectra model does not directly use the Base features, and yet outperforms the Base model based only on the photometry contained in the SED. This shows that the ML model is able to extract more information from the SED than what is simply encompassed in the estimation of effective temperature and luminosity alone, relying instead on an excess or deficit of flux at specific wavelengths when compared to the model. This shows the value obtained by using photometric data as direct input to the ML model.

Up to now, we only discussed prediction skill aggregated over all star types. Figure~\ref{fig:main-res}b-f show precision, recall and F1-score (the harmonic mean of precision and recall), split up by feature package and star type.

The skill of the classifiers greatly varies between different types of stars. For example, in terms of F1 score, the Base model works best for double stars (**), low-mass stars (lm*) and variable stars, (v*) but has near zero skill for main-sequence stars (ms*).

At the opposite end of feature inclusion, the Full model has quite high skill for low-mass stars (lm*, F1 score 0.81, recall 0.9), but the skill remains low for main-sequence stars and massive stars (ma*, F1 $\leq$ 0.34). 

Considering the individual classes, the poor precision and recall for the main-sequence stars is surprising, since the main sequence is defined on the Hertzsprung--Russell (H-R) diagram on which the Base model is based. This likely indicates poor training due to the lack of main-sequence stars classified as such at SIMBAD.

By comparison, low-mass stars (which are better populated at SIMBAD) invariably occupy the cooler, fainter parts of the H--R diagram, thus it is natural that the Base model should pick these up with comparatively high precision and recall. By contrast, massive stars are well-populated at SIMBAD (despite being rarer, they are intrinsically brighter, so easier to identify). They should invariably occupy the brighter and normally warmer parts of the H--R diagram, so we might expect the Base model to identify these well, yet it spectacularly fails to do so: including more features improves the precision and recall slightly, but they remain the class with the poorest skill. The reasons for this remain unclear.

Young stellar objects (y*o) also occupy a specific part of the H-R diagram, in the sparsely occupied region with cool temperatures but moderate luminosities. The Base package scores for this class are only moderately effective and the Spectra package makes little improvement. The Bias package returns a much better recall, likely due to the fact that most of these stars are found in an narrow strip in the plane of our Galaxy, and a few star-forming regions just away from the plane. This feeds into a strong recall and moderately high precision in the Full package. The precision may be impacted by the overlap between young stellar objects and variable stars.

Evolved stars (ev*) similarly occupy the cool, luminous part of the H--R diagram. Consequently, it is surprising that the Base package does not fit these with higher recall. The inclusion of the Spectra package is more effective at identifying them, possibly because they tend to show excess flux at infrared wavelengths. The low precision may again be impacted by the overlap between evolved and variable stars.

Considering the other well-fit classes, only some classes of double stars and variable stars should be identifiable on the H-R diagram, and then with low confidence. Consequently, it is more surprising that the Base combination of temperature and luminosity is able to return these well. Therefore, the high recall (82\%) for double stars is particularly surprising, and it is interesting to see that this recall score decreases when more features are added.

Chemically peculiar stars (pe*) and emission-line stars (em*) have lower skill in the Base package. These kinds of stars do preferentially occupy parts of the H--R diagram, but are very heterogeneous classes that cover a wide range of physical characteristics, and normally require detailed spectral observation to classify. This breadth of properties translates into a generally poor fit across the different feature packages, as there are multiple potential characteristics in feature space that a ML analysis could declare significant, while other valid parts of feature space could be excluded.

\subsection*{``Extended'' dataset}

We now proceed to our second dataset, which extends the main dataset with an additional 80\,000 main-sequence stars from random fields on the sky. We expect the addition of more main-sequence stars, the most-populous type of stars in our Galaxy but the least sampled in SIMBAD, to improve the skill of our classifier. We only chose to extend this class, due to the abundance of main sequence star data which is largely left unlabelled due to being sufficiently common.

\begin{figure}[h!]
	\centering
a)\includegraphics[width=0.45\textwidth]{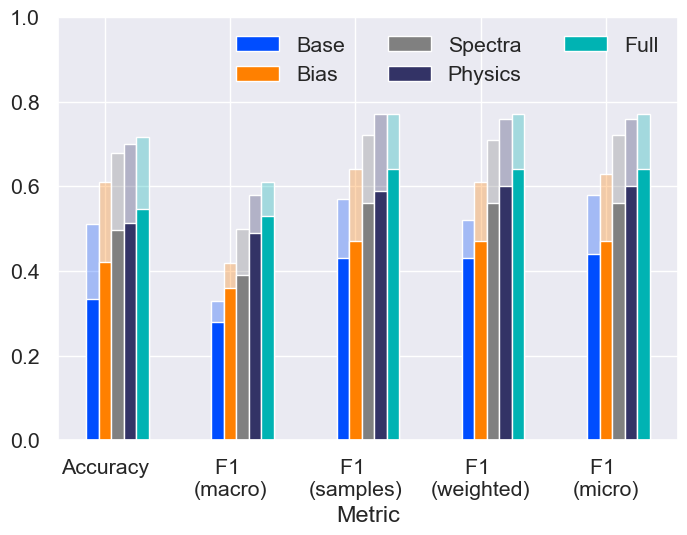}b)\includegraphics[width=0.45\textwidth]{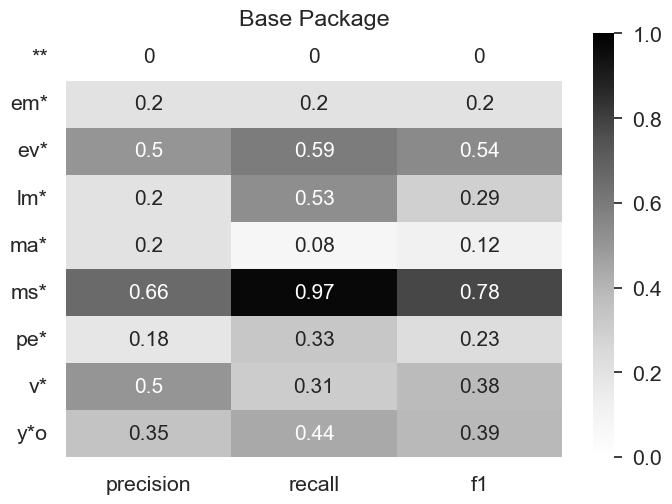}
c)\includegraphics[width=0.45\textwidth] 
{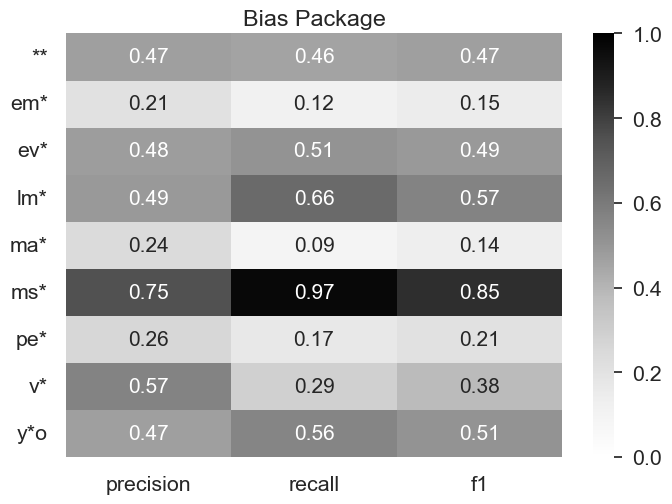}d)\includegraphics[width=0.45\textwidth]
{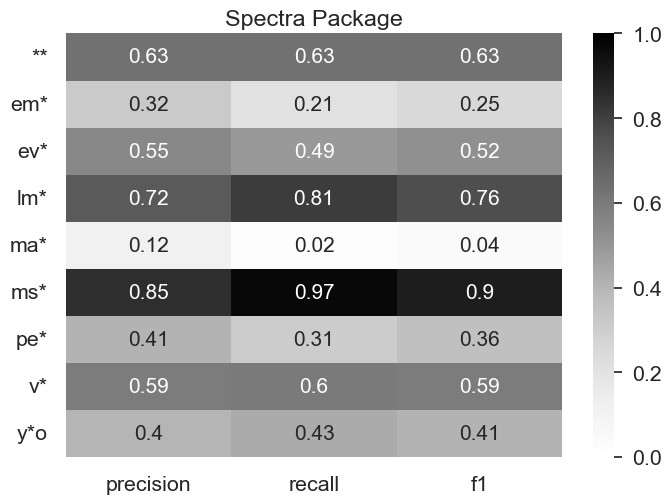} e)\includegraphics[width=0.45\textwidth]{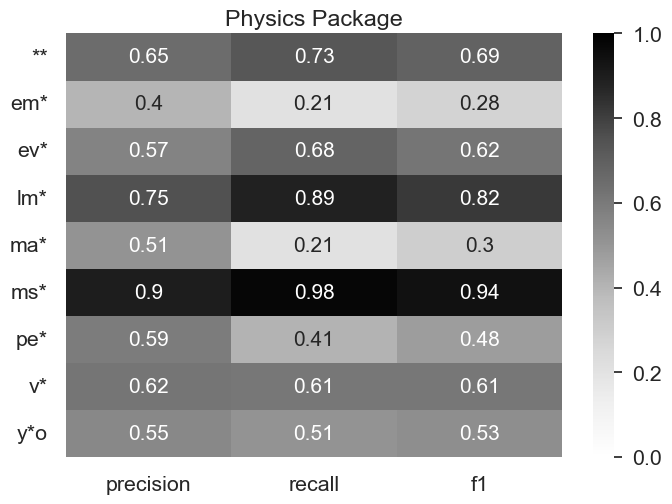}f)\includegraphics[width=0.45\textwidth]{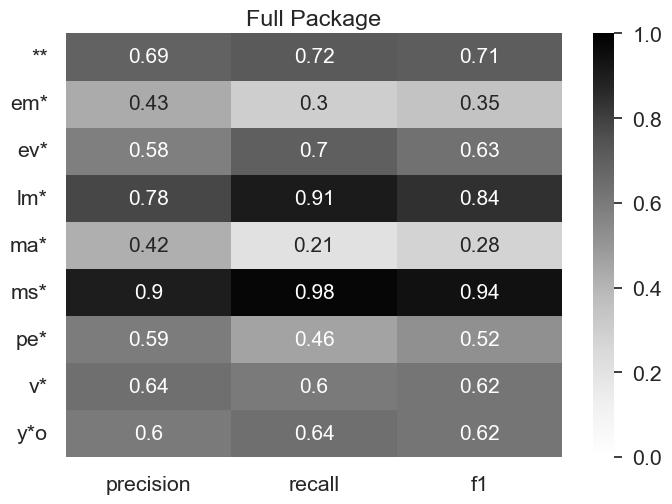}
    \centering
    \begin{tabular}{c|c|c|c|c|c|c|c|c|c}
        \hline
        \multicolumn{10}{c}{Test set sample counts per class.}\\
        \hline\hline
         & ** & em* & ev* & lm* & ma* & ms* & pe* & v* & y*o \\
        \hline\hline
        Samples & 5300 & 464 & 6484 & 1168 & 102 & 17028 & 466 & 9438 & 570\\
        \hline
    \end{tabular}
	\caption{\label{fig:augmented-res} Accuracy on the extended dataset (main dataset plus additional main-sequence stars) and comparison to main dataset, for models trained with the five different feature packages. Panel a) shows the results aggregated over all star types. The solid bars are the skill with the models trained and tested on the main dataset (same as Fig.~\ref{fig:main-res}a), the shaded bars are the skill of the models trained and tested on the extended dataset. Panels (b-f) show the results for individual star types with the extended dataset.}

\end{figure}

The results are shown in Fig.~\ref{fig:augmented-res}. Consistently across all feature packages and for all unified metrics, the models trained and tested on the extended data have higher skill than their counterparts on the main dataset (Fig.~\ref{fig:augmented-res}a); the increase in accuracy is of order 0.1-0.2, which is an increase of ~30-50\% with respect to the accuracy on the main dataset.

When considering main-sequence stars as an individual class, the differences are particularly stark: the recall on the Base model increases from 0.00 in the main dataset to 0.97 in the extended dataset, indicating more main-sequence stars are being correctly identified. This appears to be mostly at the expense of correctly identifying double stars, where the recall decreases from 0.82 to 0.00, highlighting the overlap in these two types of objects on the H--R diagram. Including the SED fits in the Spectra package improves the fitting of double stars again so that both double and main-sequence stars receive a healthy F1 score in all packages including them (0.63--0.71 and 0.90--0.94, respectively). This result shows that visible traits in the SED can be exploited by the classifier, as the SIMBAD database is dominated by spectroscopic, eclipsing and various types of interacting binaries, rather than component separation in visual binaries.

Overall, the addition of the main-sequence stars does not have a significant negative impact on the F1 scores of any other class, with the exception of some classes in the Base package. In general, the recall remains similar for the other classes, indicating few stars from other classes are now being misclassified as the majority main-sequence class. The precision scores generally show an increase too, particularly for the massive-star class, indicating fewer stars in the main-sequence class are now being misclassified as well, and it is this increase in precision for massive stars, and the increase in both precision and recall for main-sequence stars, that are main drivers of the increased F1 scores for the extended dataset. This shows that this classification problem is still being limited by a lack of adequate training data for specific sub-classes of objects.

Inspired by this result,  
we designed a final experiment to get more insights on how decreasing or increasing the number of training samples impacts the performance of the classifier. For this, we started with the main dataset and systematically reduced the training samples in 5\% steps individually for each star type. For example, we use the main dataset but with only 95\%, 90\%, 85\%, ..., 0\% of variable stars. We repeat this process for each star type in turn, resulting in 171 datasets across all types. For each dataset, we train the Full feature package model and evaluate it on the test dataset, which remains unchanged. To avoid impacting the other classes, if a star received two labels (e.g., evolved and variable star), we only removed the entry from the class undergoing reduction in that dataset. 

\begin{figure}
	\centering
	a)\includegraphics[width=0.45\textwidth]{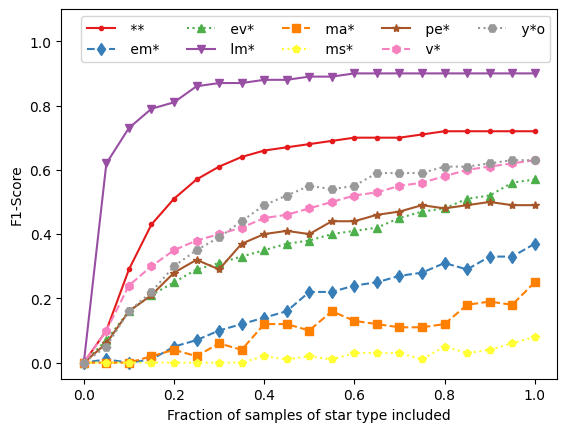}b)\includegraphics[width=0.45\textwidth]{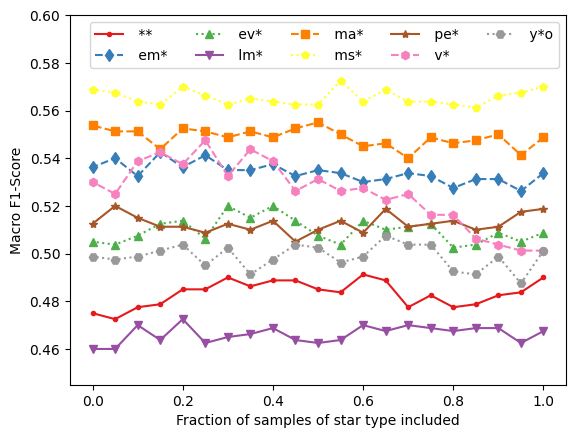}
	\caption{\label{fig:add-exp}Results from experiment systematically reducing the training samples of individual star types. a) F1 score of the star type systematically reduced, b) macro-averaged F1 score over all star types except the systematically reduced one. Details see text. }
\end{figure}

If we were to compute the macro F1 score for each of those experiments, we would conflate two different effects: (1) the change of skill on the star type whose samples are reduced and (2) the effect on overall skill if fewer samples are available. 
In order to disentangle those effects, we use two different F1 scores: first, one time we compute the F1 score solely of the star type that is systematically reduced; second, we use the macro F1 score over the other eight star types. These are respectively shown in Fig.~\ref{fig:add-exp}, panels (a) and (b). Panel (a) shows the clear (and unsurprising) result that, for all star types, the skill is reduced as fewer training samples are included, finally reaching zero. However, the rate of this decay could not have been known a-priori without this analysis. This rate is informative:  if the curve is very flat, then this indicates that the training algorithm has reached a plateau - in which case including more samples would likely not improve performance further. On the other hand, if performance in the experiment is gradually increasing with number of included samples, then this indicates that including more samples might likely lead to a further increase in performance.

We see that effect of removing samples very much depends on the type of star: the relatively flat curves for low-mass stars and double stars indicate that we are likely at the maximum F1 score achievable for these stellar types, while the F1 score for the other types is still increasing towards the right of the graph. Due to the computational time required, this training set only contains 10\% of the stars classified in SIMBAD, however these results suggest that higher F1 scores could be obtained by analysing an increased number of objects from the SIMBAD database, and especially by including more objects from these other classes.

Gradually removing a class in this fashion seems to have little effect on the skill of the other star types (panel b). One possible exception are variable stars. Here the results suggest that reducing their number in the training set actually slightly \textit{increases} the skill on the other star types. This is not surprising, as variable stars tend to overlap with other star types. The skill increase may suggest that if the ML algorithm learns to identify them then the algorithm may consequently also mislabel similar star types as variable stars, presuming guilt by association. An explainable AI study would be needed to confirm this suggestion.

The result that decreasing one star type in the training has little impact on skill for the other star types is in congruence from the results of the experiment with the extended data, where the additional main-sequence stars greatly increased the skill on the main-sequence stars themselves, but not for the other star types. The overall increase in accuracy for that experiment stems from the fact that in the extended data, main-sequence stars are the majority class and thus have a large impact on overall skill. 

Finally, as means of better understanding the final classifier’s prediction behaviour, specifically pertaining to the treatment of missing values in XGBoost, we analysed the SHAP importance values for our test dataset.

\begin{figure}
    \centering
    \includegraphics[width=0.67\linewidth]{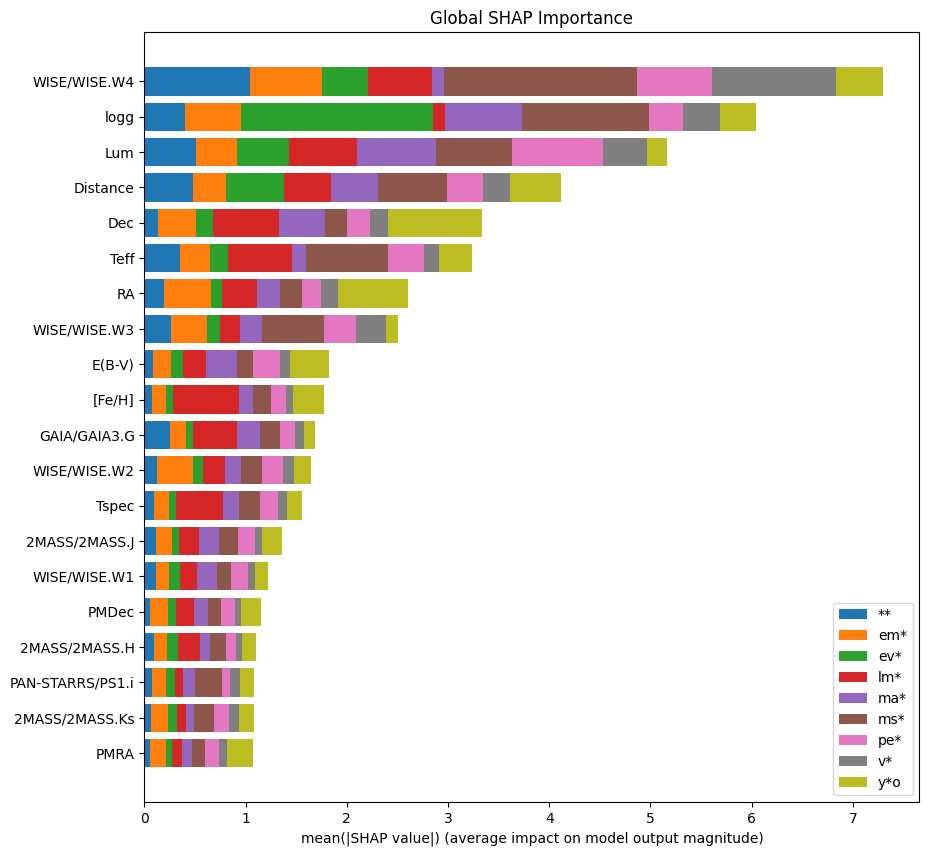}
    \caption{Top 20 feautures of the full feature set ranked by their mean absolute SHAP importance values}
    \label{fig:shap_full}
\end{figure}

We utilized the python library SHAP created by \citet{SHAP}, to extract approximations of Shapley values from coalitional game theory, which measure the contribution of each feature on the final model's prediction. These local explainability measures can in turn be aggregated across samples, giving rise to a global explainability metric, for which the results on the test set with the full feature package can be seen in Fig.~\ref{fig:shap_full}.
Several findings arise from this plot, most notably in regards to the high amount of missing values in the dataset, that XGBoost seems to quite successfully learn the pattern of missingness, supporting the notion that the missing values are indeed “missing not at random”. A common assumption is that more densely packed features contain more information, which at first glance may seem to be the case as 17 out of the top 20 most abundant features are contained in this list, however the most relevant feature to the model is the WISE.W4 band.

From a data-centric perspective, this is highly surprising as it is among the less densely packed features available. This implies that XGBoost, indeed learns the missing value pattern, and in turn utilizes such realizations in the final model's predictions. An example of this can be seen in the class most impacted by this band, which upon closer analysis of the local explainability, reveals that out of the top 50\% most impacted main-sequence stars, 95\% were missing values. Similarly, the top 5\% of samples impacted most by this band, had over 92\% missingness.
This leads us to two significant data-centric takeaways, namely that missing value patterns are truly learned to act as discriminative features by the model’s default pathway choices and that, as is to be expected, the impact of bands and features varies between star types.

From an astrophysical perspective, this is marginally less surprising. Temperature and luminosity (and, by implication, log($g$), are some of the primary parameters affected by properties such as stellar mass and evolutionary stage, so are very effective discriminants of stellar class to human classifiers. Similarly, distribution on the sky in RA and Dec can be used as a secondary classifier: e.g., young stellar objects will always cluster in star-forming regions and (together with massive stars) will be found almost uniquely in the plane of the Milky Way; low-mass stars will not be found at large distances, etc. The high importance of mid-infrared data, particularly from the \emph{WISE} satellite, may be attributable to the presence or absence of excess infrared flux, as found among young stellar objects, evolved stars, and some variable and massive stars. \emph{WISE} fluxes in excess of the fitted stellar atmosphere model can act as discriminators to positively identify these classes, hence also as evidence against the other classes. This extends to presence or absence of these data which, especially for \emph{WISE} band 4, are limited by sensitivity: absence of \emph{WISE} observations appears to be being used by the classifier as a proxy for lack of excess infrared flux.

\section*{Conclusions}
In this work, we have trained machine-learning classifiers for classifying stars. Photometric data from VizieR and higher-level data products from the PySSED spectral-energy-distribution fitting algorithm were used as inputs, for a training dataset comprising 10\% of stars classified in SIMBAD as belonging to one or more of nine categories of stars. The resulting classifier has an accuracy of $\sim$0.5 when using all available input variables. When adding additional 80\,000 main-sequence stars to combat the severe class imbalance in the training data, the accuracy increased to $\sim$0.7.
In additional experiments, we disentangled the contribution of individual groups of input variables. Finally, we conducted an experiment with systematically decreasing the numbers of individual star types, and found the prevalence of a particular star type in the training set greatly influences the skill of the model on that particular star type, but has little to no effect on the skill for the other star types.

This work is a first attempt to use supervised ML techniques for stellar classification on stars from the SIMBAD database. Key challenges are the high sparsity of the data -- meaning there are many missing values -- and the high class imbalance, as the number of available samples varies massively between the different star types. While the skill of the best model is not yet high enough to be usable as a reliable tool in practice, this work shows the principal feasibility of using ML for such a problem.

We suggest several avenues for further work: 
(1) Using more data on under-represented star types. Our results suggest that adding more data can increase the predictive skill at least for some star types. 
(2) Using a different taxonomy of star types. Several of the classes we adopted from SIMBAD are very heterogeneous (e.g., peculiar stars, emission line stars, variable stars and main-sequence stars). A larger number of taxonomic classes that are more cleanly separated in their physical properties may both reduce the problems of multi-label classification and provide greater utility to the astrophysical community. 
(3) Developing and using physics-informed methods for data imputation. In our work we solely used XGBoost as ML algorithm because it can natively deal with missing data. Initial tests with simple data-imputation methods were unconvincing. However, it would be interesting to develop data imputation methods specialized for our use case. Specifically, methods that incorporate \emph{a priori} physical knowledge into statistical and ML methods might be promising, i.e., physics-informed methods \cite{karniadakis2021physics}.
(4) Lastly, we chose these 5 distinct feature packages with the goal of evaluating the bias in the data and the importance of the SED fitting procedure, however believe that this experiment can be further extended to the generation of different, more numerous feature packages, examining the impact of various variables, such as for example a physics feature set without the Base variables (Temperature and Luminosity).

Finding a good way of filling up the large number of gaps in stellar SED data, e.g., by use of the stellar atmosphere models used in PySSED, would remove one of our main challenges, namely the high sparsity of the data. Not only is non-sparse data often easier to separate on a fundamental level, in practice it also allows using a much wider variety of ML techniques.

\section*{Appendix}

\begin{figure}
\centering
a)\includegraphics[width=0.3\textwidth]{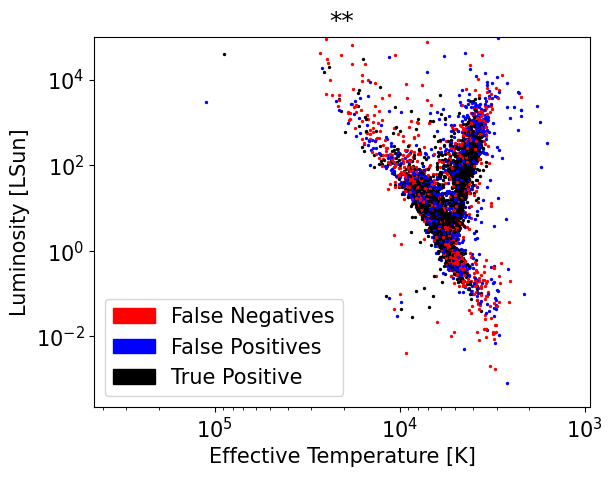}
b)\includegraphics[width=0.3\textwidth]{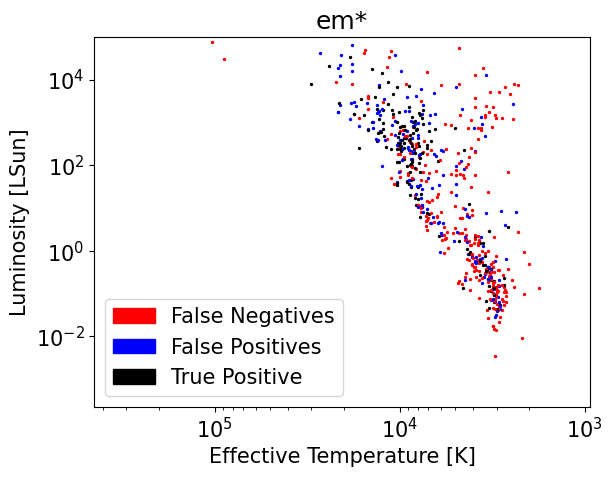}
c)\includegraphics[width=0.3\textwidth]{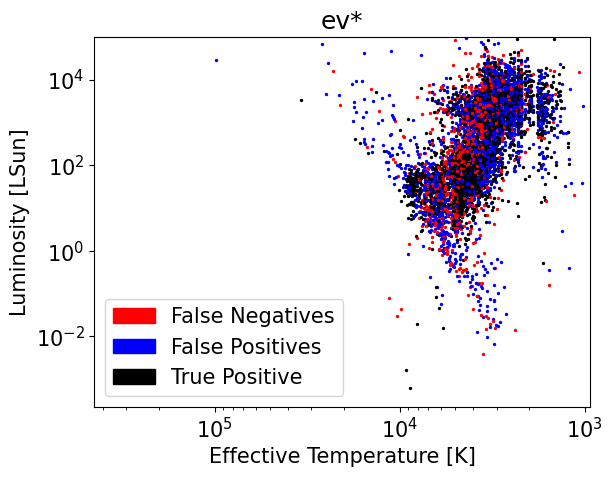}
d)\includegraphics[width=0.3\textwidth]{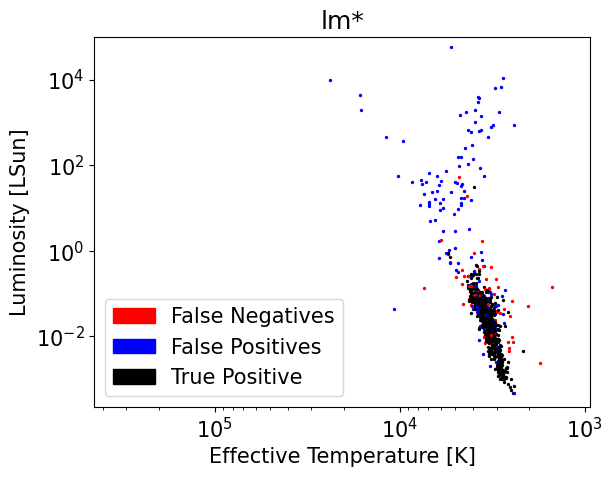}
e)\includegraphics[width=0.3\textwidth]{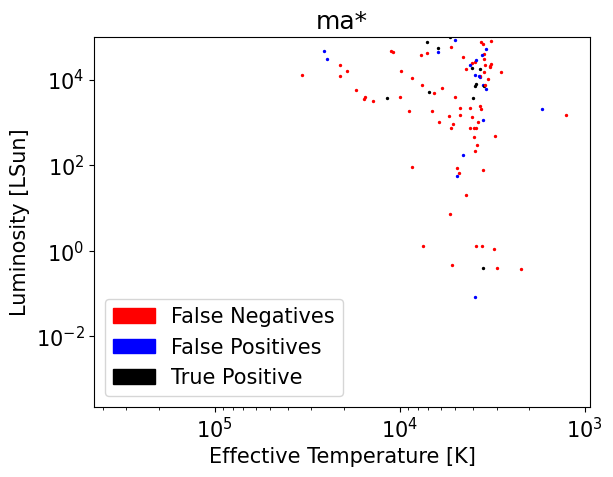}
f)\includegraphics[width=0.3\textwidth]{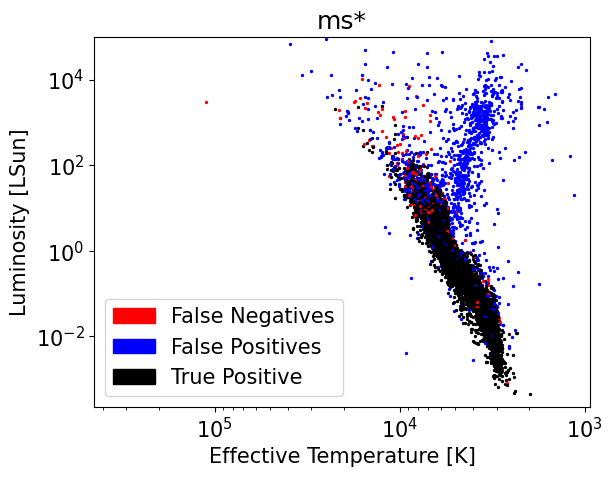}
g)\includegraphics[width=0.3\textwidth]{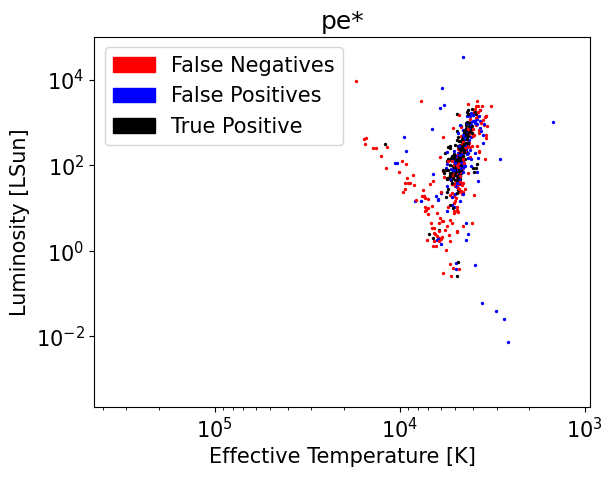}
h)\includegraphics[width=0.3\textwidth]{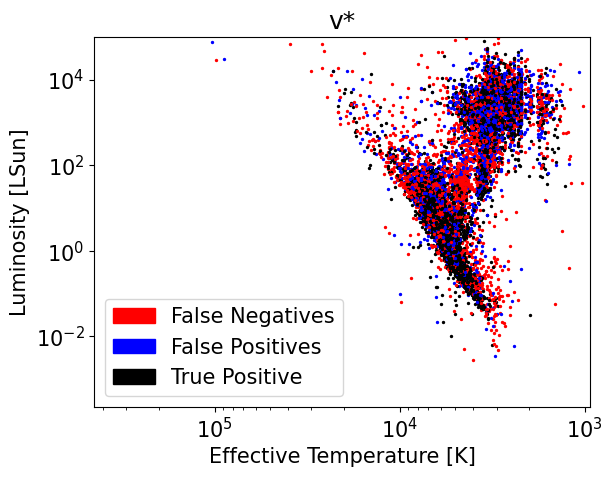}
i)\includegraphics[width=0.3\textwidth]{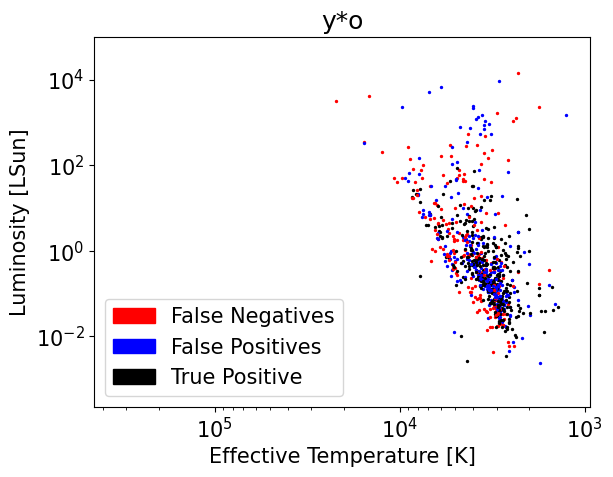}
 \caption{Extended Dataset Classification Results Visualised in HR Diagrams}
\end{figure}

\begin{figure}
\centering
a)\includegraphics[width=0.35\textwidth]{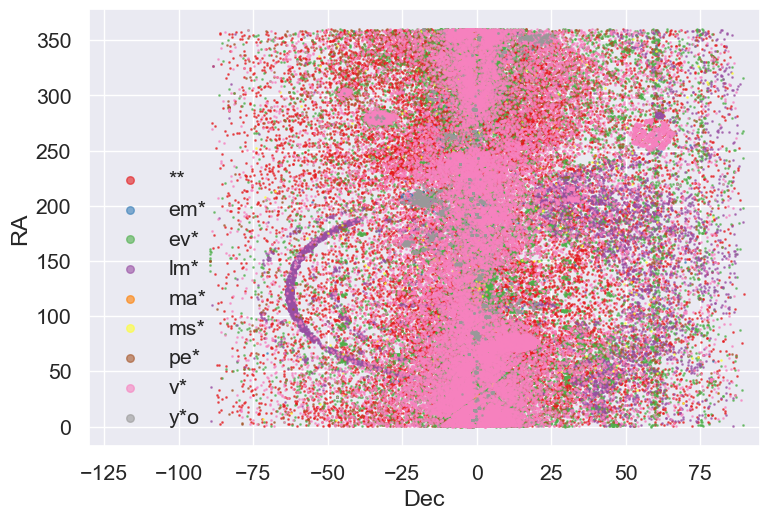}
b)\includegraphics[width=0.35\textwidth]{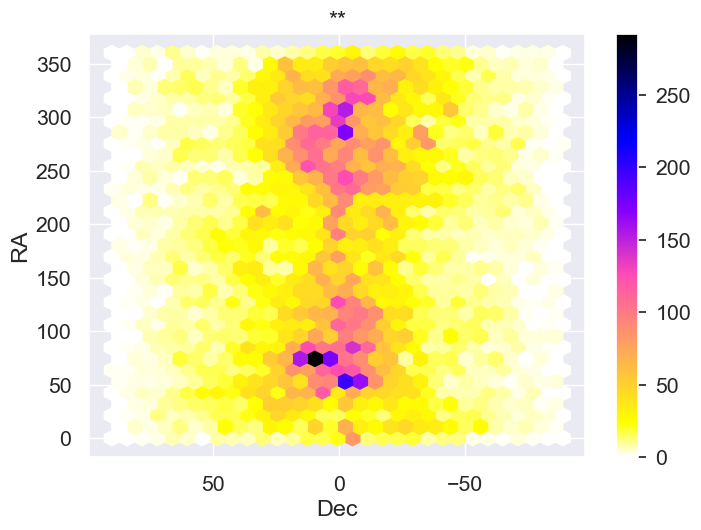}
c)\includegraphics[width=0.35\textwidth]{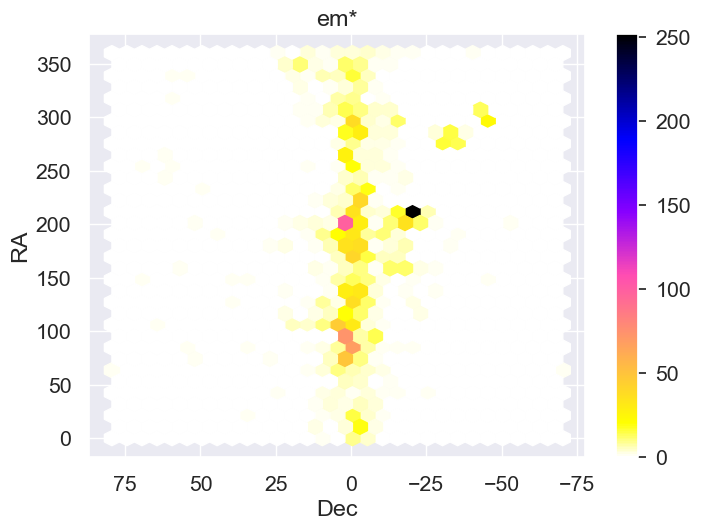}
d)\includegraphics[width=0.35\textwidth]{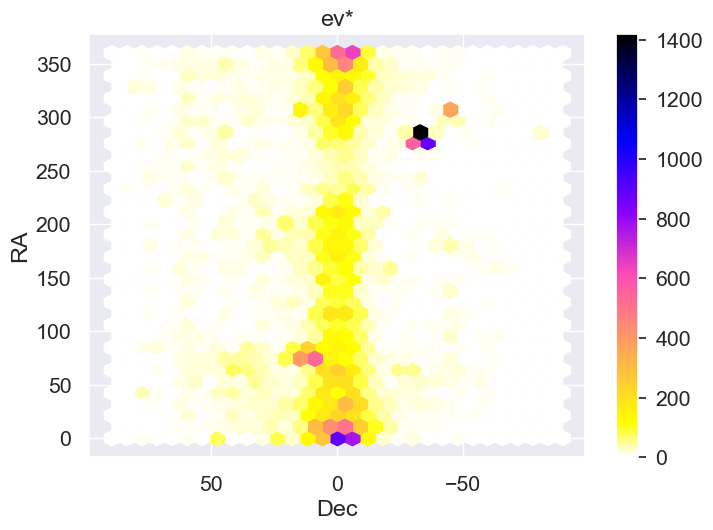}
e)\includegraphics[width=0.35\textwidth]{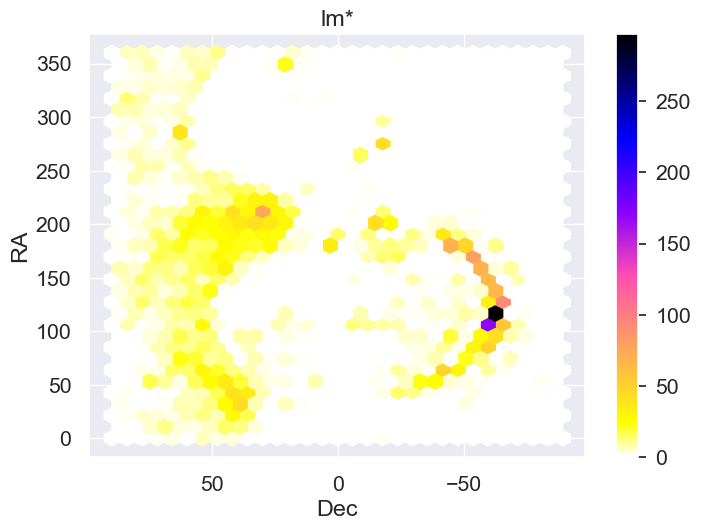}
f)\includegraphics[width=0.35\textwidth]{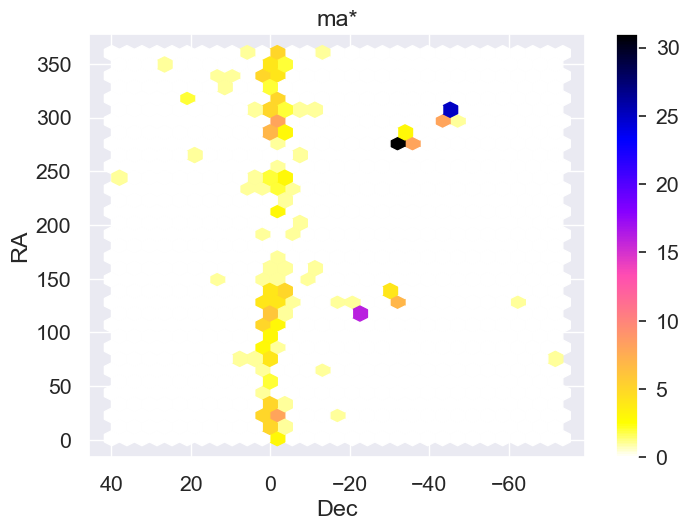}
g)\includegraphics[width=0.35\textwidth]{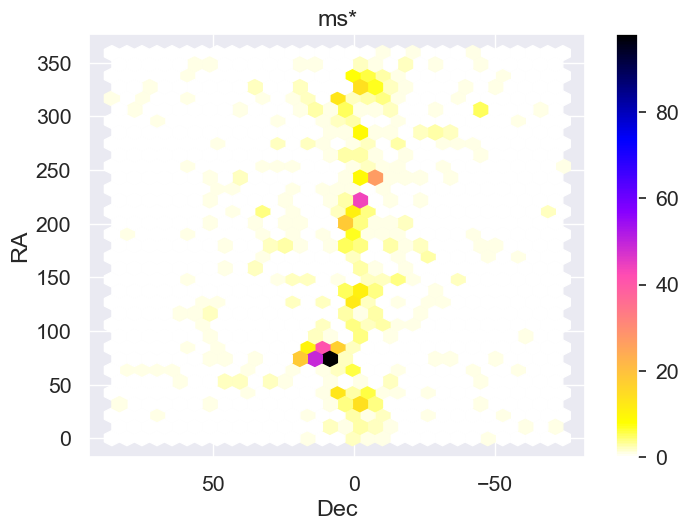}
h)\includegraphics[width=0.35\textwidth]{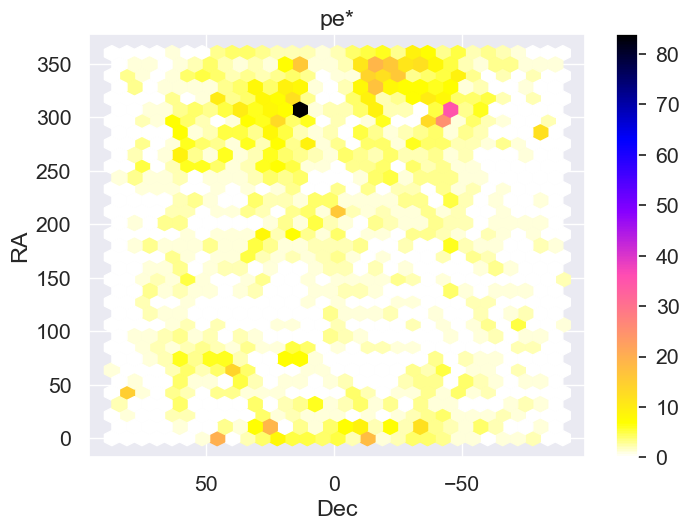}
i)\includegraphics[width=0.35\textwidth]{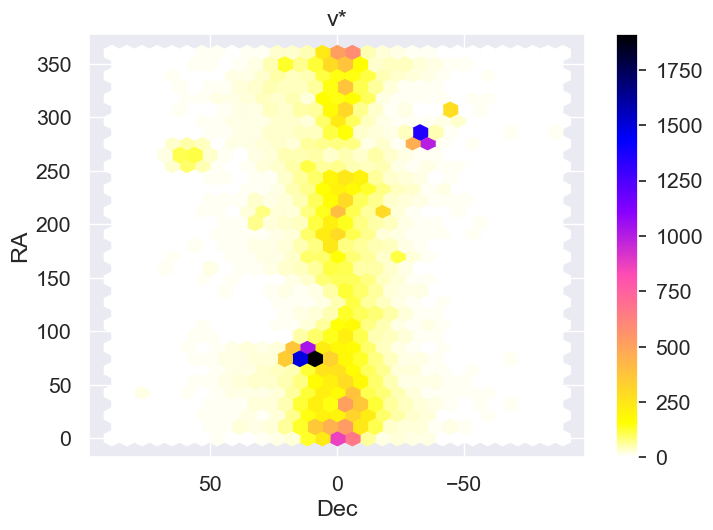}
j)\includegraphics[width=0.35\textwidth]{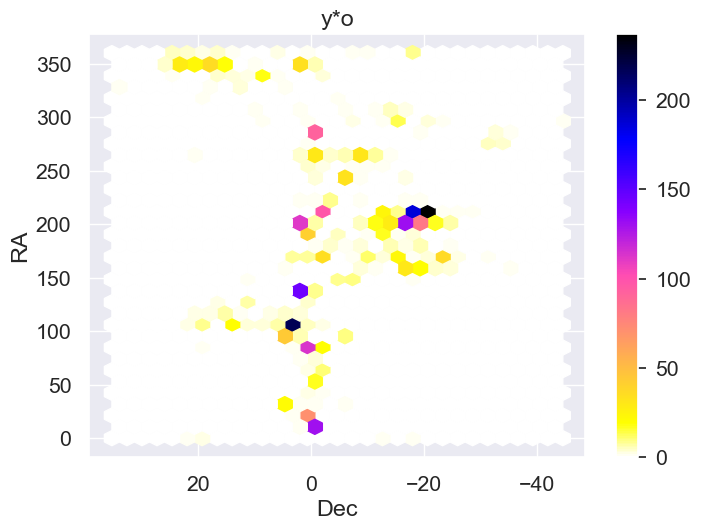}
 \caption{a) star locations of the main dataset b)-j) Density plots of individual star locations in the main dataset.}
\end{figure}

\section*{Data and software availability} 


\subsection*{Source data}

Source data for this work comes from the VizieR database (\url{https://vizier.cds.unistra.fr/viz-bin/VizieR}) and is analysed using the PySSED pipeline \cite{mcdonaldpyssed}. Classification data comes from the SIMBAD database (\url{https://simbad.u-strasbg.fr/simbad/}).

\subsection*{Underlying data}
The data used in this study is a subset of the SIMBAD dataset, extended with additional data from the PySSED Software. The complete dataset used for this study can be accessed at https://zenodo.org/records/10377253.

The data is stored as .csv files.

\begin{quote}
\ \\
This project contains the following underlying data:
\begin{itemize}
	\item 10percent\_full.dat. (Main Dataset Data)
	\item 10percent\_labels.txt. (Main Dataset Labels)
    \item df\_plus\_ms.dat. (Extended Dataset Data)
    \item classes\_plus\_ms.dat (Extended Dataset Labels)
\end{itemize}

Data are available under the terms of the \href{https://creativecommons.org/licenses/by/4.0/}{Creative Commons Attribution 4.0 International} data waiver.
\end{quote}






\subsection*{Software availability}

\begin{itemize}
    \item Stellar Classification Code (\url{https://github.com/explore-platform/s-phot_stellar_classifier}),\\ 
    Archived at the time of publication: \url{https://doi.org/10.5281/zenodo.10377253}.\\ Licensed under \href{https://creativecommons.org/licenses/by/4.0/}{Creative Commons Attribution 4.0 International}
    \item PySSED Code v0.3 (\url{https://github.com/explore-platform/pyssed_explore/tree/v0.3})
\end{itemize}



\section*{Acknowledgements}
We thank Roman Kern, Vedran Sabol and Manuela Rauch for their inputs and discussions.


{\small\bibliographystyle{unsrtnat}
\bibliography{bibliography, bibliography_emma}}




\end{document}